\documentclass[10pt,twocolumn,letterpaper]{article}

\usepackage{cvpr}      

\usepackage{graphicx}
\usepackage{amsmath}
\usepackage{amssymb}
\usepackage{booktabs}
\usepackage{placeins}
\usepackage{xcolor}

\usepackage{array}

\newcolumntype{C}[1]{>{\centering\let\newline\\\arraybackslash\hspace{0pt}}m{#1}}

\newcommand{\bluet}[1]{\textbf{\color{blue}#1}}
\newcommand{\redt}[1]{\textbf{\color{red}#1}}

\newcommand{\net}{\ensuremath{\mathcal{N}}}

\usepackage[pagebackref,breaklinks,colorlinks]{hyperref}

\usepackage[capitalize]{cleveref}
\crefname{section}{Sec.}{Secs.}
\Crefname{section}{Section}{Sections}
\Crefname{table}{Table}{Tables}
\crefname{table}{Tab.}{Tabs.}

\begin{document}

\title{Unsupervised HDR Imaging: What Can Be Learned from a Single 8-bit Video?}

\author{Francesco Banterle\\
ISTI-CNR\\
{\tt\small francesco.banterle@isti.cnr.it}
\and
Demetris Marnerides\\
Independent Researcher, UK\\
\and
Kurt Debattista\\
Warwick University, UK
\and
Thomas Bashford-Rogers\\
University of West England, UK
}

\maketitle

\begin{abstract}
Recently, Deep Learning-based methods for inverse tone-mapping standard dynamic range (SDR) images to obtain high dynamic range (HDR) images have become very popular. These methods manage to fill over-exposed areas convincingly both in terms of details and dynamic range. Typically, these methods, to be effective, need to learn from large datasets and to transfer this knowledge to the network weights. In this work, we tackle this problem from a completely different perspective. What can we learn from a single SDR video? With the presented zero-shot approach, we show that, in many cases, a single SDR video is sufficient to be able to generate an HDR video of the same quality or better than other state-of-the-art methods. 
\end{abstract}

\section{Introduction}

To capture the full range of color and shades of brightness in the real world, high dynamic range (HDR) imaging is employed. Even though modern sensors, cameras, and smartphones can capture HDR imagery, a large amount of content was and still is captured in standard dynamic range (SDR) or is converted to SDR after capture. 

When presenting this content on HDR displays \cite{Seetzen+2004}, or using this imagery for applications where HDR values are required \cite{Debevec+2002}, SDR values need to be boosted to HDR; a process known as Inverse Tone Mapping \cite{Banterle+2006}.

Researchers have proposed a wide variety of approaches to solving this problem, from straightforward linear functions \cite{Akyuz+2007} to, more recently, deep-learning (DL) based solutions \cite{Eilertsen+2017,Endo+2017}. Typically, these DL approaches outperform the existing methods and are mostly based on training a convolutional neural network (CNN) to encode a mapping from SDR to HDR. To achieve this, a large set of SDR/tone mapped and reference HDR image pairs is required to train a general mapping.    


We propose a fundamentally different approach based on the observation that much of the information required for inverse tone mapping may be present in an SDR video sequence. 
This can be a result of a variety of effects that are present in videos but not in still images. For example, motion in the scene or from the camera can uncover detail that was badly exposed in earlier frames. In addition, changes in the lighting of the scene, or luminance variations due to automatic exposures from the camera can also create a similar effect, where information otherwise lost in some frames exists in some others.

\begin{figure*}[ht]
    \centering
    \includegraphics[width=\linewidth]{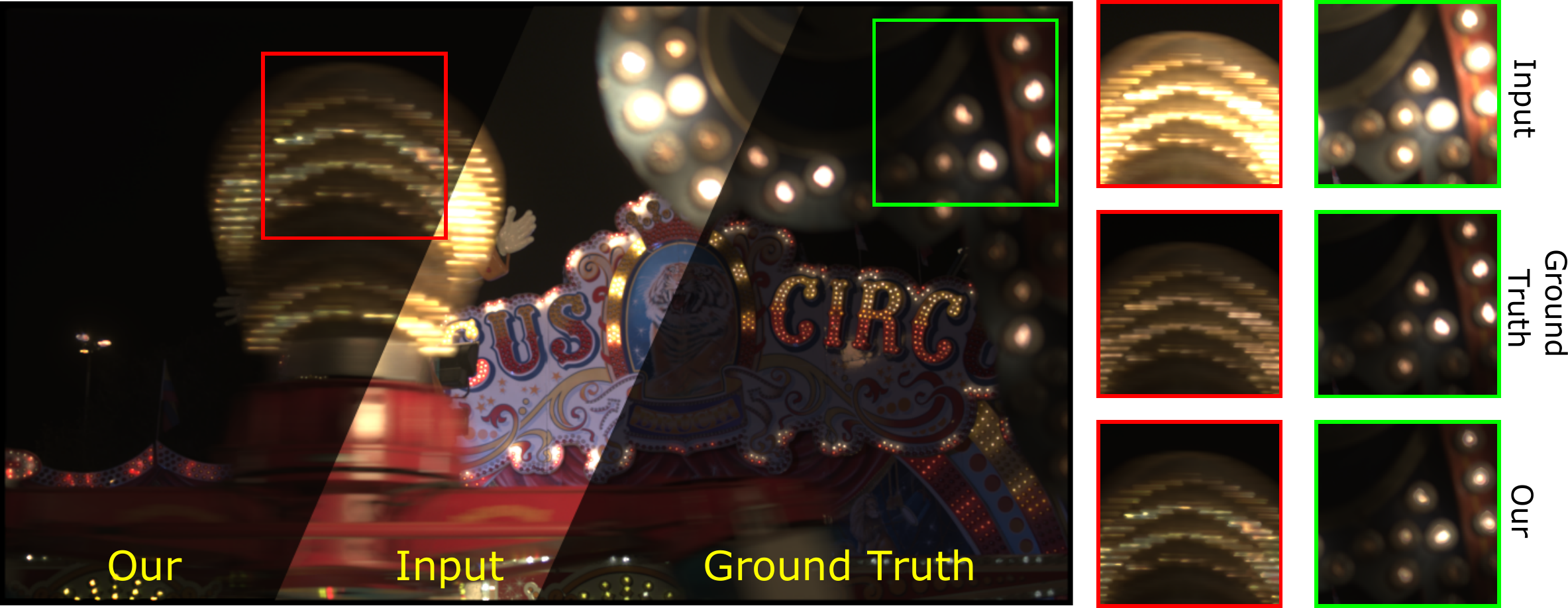}
    \caption{\small{An example of our inverse tone mapping operator applied to an SDR version frame from the Carousel fireworks 02 sequence [13]. Our method can recover missing texture, colors, and dynamic range details in a convincing way.}}
    \label{fig:teaser}
\end{figure*}

Our approach attempts to gather and distill this information present in a single SDR video in order to recover information in over-exposed and under-exposed areas of the same video. Figure \ref{fig:teaser} shows results of our method. We define a new pipeline for expanding the dynamic range of SDR content using deep-learning approaches. This optimization relies only on the frames of the SDR video that is processed in a zero-shot fashion. In the presence of only a single SDR video, there is no ground truth HDR data for training and the method uncovers HDR patterns embedded in the underlying SDR signal using self-supervision. The neural network weights that hold all the knowledge for inverse tone mapping are uniquely learned for each video without relying on external datasets of HDR images or other videos, which are still limited in quantity \cite{Santos+2020}. 



In summary, we propose a novel inverse tone mapping operator (ITMO) for expanding SDR videos that uses a zero-shot strategy and self-supervision. Our approach, even though unsupervised, broadly outperforms state-of-the-art fully supervised ITMO methods both visually and across several metrics.  Our main contributions are:
\begin{itemize}
    \item A zero-shot solution that exploits exposure information present in SDR videos to reconstruct HDR sequences in a self-supervised manner;
    \item An unsupervised, straightforward, and effective architecture for expanding SDR videos to HDR without the need of a comprehensive dataset. 
\end{itemize}

The source code of this work will be made available online.


\section{Related Work}
ITMOs generate an HDR image/video from an original SDR version that is quantized at 8-bits \cite{Banterle+2006}. This problem is ill-posed because there is not much information left in under-exposed and over-exposed areas.

\subsection{Classic Methods}
ITMOs, not employing deep learning, can be classified into three main classes: global, local, and user-based. On one hand, global ITMOs define an expansion function that gathers global statistics from the image and applies it to all pixels. These ITMOs can use linear functions \cite{Akyuz+2007}, multi-linear functions \cite{Meylan+2006} and gamma functions \cite{Landis2002,Masia+2009, Masia+2017, Bist+2017}. On the other hand, local ITMOs define an expansion function that varies per pixel locally exploiting both local and global statistics from the image. Several strategies have been proposed. For example, some operators generate an expand map (a spatially varying function) for guiding the expansion only in certain area of high luminance \cite{Banterle+2006,Rempel+2007,Huo+2014,Kovaleski+2014}.
In user-based methods, the user drives the expansion and details recovery. For example, Wang et al. \cite{Wang+2007} proposed a solution in which a user recovers the dynamic range and details of an SDR image using clone-tools and inpainting techniques similar to modern image editors. Another example of such methods is Didyk et al.'s work \cite{Didyk+2008}. In this work, a semi-automatic classification interface allows users to classify pixels into area consisting of diffuse, reflections, and light sources. Then, only reflections and light sources are expanded by applying an adaptive non-linear function.

\subsection{Deep Learning-based Methods}
Recently, several ITMOs have been proposed using different DL architectures.
DL-based methods have largely taken two approaches. The first is to directly reconstruct an HDR image from SDR and the second predicts a set of SDR exposures which are fused to generate an HDR image \cite{debevec1997recovering}. Eilertsen et al. \cite{Eilertsen+2017} masked out well-exposed regions which were reconstructed by a linear operator, and overexposed regions which were reconstructed by a UNet. Eilerstein et al. \cite{Eilertsen+2019} extended this work for temporal stability via training regularization. Marnerides et al. \cite{Marnerides+2018} used a multi-branch network to directly reconstruct the HDR image where each branch was designed to capture different features for reconstruction. Approaches have also been proposed to reverse the camera pipeline to synthesize HDR images, for example, Yang et al. \cite{yang2018image} also used a UNet and Liu et al. \cite{liu2020single} reconstruct images using a series of networks. Santos et al. \cite{Santos+2020} proposed an ITMO based on pretraining a network for inpainting then specializing this network for ITMO based on masking. Endo et al. \cite{Endo+2017} was the first work to predict a set of exposures via an autoencoder that are then fused to generate a HDR image. The creation of multiple exposures is similar to our work, except their method relied on a large set of training images to learn the mapping from SDR to HDR. Recently, Zhang et al.\cite{Zhang+2021} showed that processing high-frequency and low-frequency parts of an image separately can improve the reconstruction process. The NTIRE 2021 Challenge on High Dynamic Range Imaging \cite{NTIRE21} presented several supervised HDR reconstruction methods which proposed a range of network architectures and datasets for the evaluation of static images, although these methods are not compared with the state-of-the-art.
In terms of video, Kim et al. \cite{kim2019deep} proposed a super-resolution and inverse tone mapping approach designed for video applications that directly produced HDR frames. They reconstruct low and high-frequency information separately and include upscaling of the high-frequency information, which are then combined into the final frame. Both dynamic range expansion and super-resolution are computed per frame without an explicit mechanism for enforcing temporal coherence.

These approaches are all based on the same underlying concept of applying transformations to a ground truth set of HDR images to synthesize an SDR dataset, then learning the mapping from SDR to HDR or a set of exposures. While providing a general approach to inverse tone mapping, these methods have drawbacks in that they cannot be specialized to a particular type of content, and require significant dataset sizes and training to learn the mapping. 

\subsection{Self-supervised Methods for Imaging}
Recently, self-supervised methods have become more popular thanks to their performance and the use of limited or no datasets. 
Shocher et al. \cite{Shocher+2018} introduced zero-shot methods for inverse imaging problems and showed that such strategies can be effective and produce convincing results. They proposed a zero-shot super-resolution method where the key idea is to create a dataset using a downsampling operator on the input image itself. Then, this tailored dataset was used to train a convolutional network; after training it was used to upscale the input image. The key observation is that the image has repetitions of details at different scales that can be exploited.
With a similar aim but a different methodology, Ulyanov et al. \cite{Ulyanov+2020} proposed the Deep Image Priors framework, where imaging problems such as denoising, inpainting, super-resolution, deblocking, etc. are solved by optimizing the network parameters exploiting a prior degradation function, $h$, that is known. In this case, no dataset is generated but $h$ (e.g., downsampling operator, blocking method, etc.) has to be defined for each problem.









\section{Self-Supervised Expansion}

The core concept behind this work is based on the observation that the exposure time when capturing SDR videos frequently changes from frame-to-frame and the same regions of the image may be recorded with different exposure times in the same sequence. This means that information about multiple exposures which can be used for inverse tone mapping is already present in many videos This indicates that training an ITMO on large datasets, as all deep learning approaches currently do, is not always required.

This motivates the design of an approach that can leverage this information for tone mapping. While a patch-based \cite{van2020high} or optical flow \cite{chen2021hdr} method could be used to find the same region of an image in different frames with different exposures, we instead use an approach based on deep learning. This is motivated by the success of deep learning for inverse tone mapping (e.g. \cite{Eilertsen+2017,Marnerides+2018,Santos+2020}) and the use of zero-shot methods with deep learning for single image operations, for example, the super-resolution approach by Shocher et al. \cite{Shocher+2018} based on a similar analysis of similar content in static images \cite{zontak2011internal}.

\subsection{Overview}
Given an SDR video as input, our method employs a CNN, \net{}, to generate additional exposures for each video frame. \net{} predicts per-pixel multiplicative residuals, $\hat{\delta}$, such that the lower exposure prediction image, $\hat{I_l}$, with an $e$ f-stop difference from the input, $I_b$, is given by $\hat{I_l} = \hat{\delta} I_b$. 
To generate a frame at higher exposure, the input frame is divided by the residual, $\hat{I_h} = I_b / \hat{\delta}$. 
This process can be repeated on the generated frames. For example, if we want to generate a -4 f-stop exposure with $e=2$, we need firstly to compute a -2 f-stop exposure $\hat{I}_{-2} = I\net{}(I)$, and then to compute our goal exposure as $\hat{I}_{-4} = \hat{I}_{-2}\net{}(\hat{I}_{-2})$.


Apart from the single video that is to be expanded, no further data needs to be used for training, making the method zero-shot. The method uses self-supervision from the SDR video as there is no ground truth HDR target. In the absence of a supervisory HDR signal for training, a tailored training dataset is generated from the SDR video that is to be expanded. 

\subsection{Tailored Dataset Generation}
\begin{figure}[t]

    \centering
    \includegraphics[width=\linewidth]{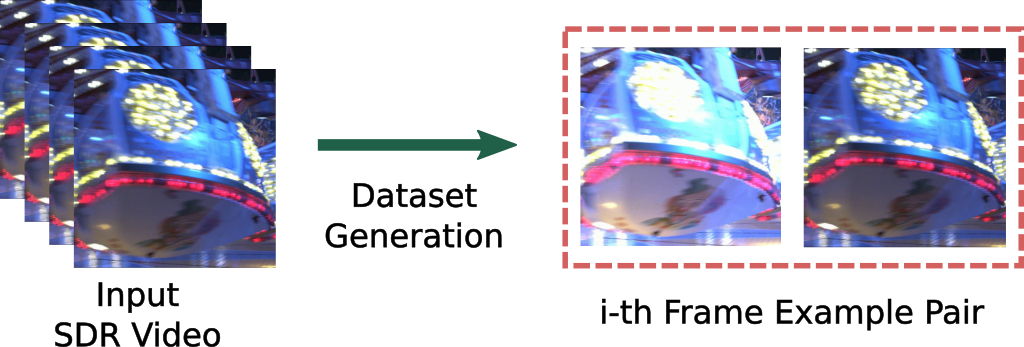}
    \caption{An example showing how the training data is generated from a video. Each SDR frame is exposed to a higher exposure. The higher exposures are then used as inputs during training, to learn the multiplicative residual mapping \net{}, using the starting SDR frames as targets.}
    \label{fig:dataset:pairs}
\end{figure}

%
The training dataset is formed by extracting a higher exposure, $I_h$, from each SDR frame, $I_b$, of the video at a base exposure value $b$. Then, $I_h$ is used to compute the multiplicative residual as: $\delta = I_b / I_h$.
Note that $I_h>I_b$, therefore $\delta \in [0,1]$.
%
At evaluation time, \net{} will instead take the original SDR frame as input, predicting $\hat{\delta}$ to compute higher and lower exposures.

Starting from an SDR input video, $V$, we assume that it is the result of exposing the ground truth HDR scene, at a base exposure value $b$. To create a training-time input for \net{}, we re-expose the frames $I_b$ to a higher exposure value $h = b+e$ forming a high exposure dataset, $V_h = \{ I_h, I_b \}$, as illustrated in Figure~\ref{fig:dataset:pairs}. The residual $\delta = I_b/I_h$ will be the target for the residual-predicting network, \net{}, during training, whereas at the inference stage, exposure $I_b$ will be the input of \net{}. The value of the exposure difference, $e$, is set to 2 f-stops; we found this value to be the largest value we could use without leading to too large over-exposed areas in the re-exposed input frame.

To ensure model robustness with respect to luminance and exposure variations, we employ a data augmentation technique, where the starting exposure $b$ is randomly shifted by a small amount $s \sim \mathcal{U}(0,0.25)$ to a higher exposure, $\tilde{b} = b + s$. The corresponding higher exposure dataset, $V_h$, is further shifted to an exposure $\tilde{h} = h+s$, forming the final dataset $\mathcal{D} = \{ V_{\tilde{b}}, V_{\tilde{h}} \}$. In our implementation, the frames are subsampled at a rate of 6 frames per second, as this was found to provide better stability when training (and it is also a common factor of the traditional frame rates of 24 and 30).

The exposure function is given by:
\begin{equation}\label{eqn:exposure}
    I_{\text{exp}} = \bigl[(g^{-1}\bigl( g(I) \cdot 2^{\Delta v} \bigr) \bigr]_0^1 = \bigl[\bigl( I \cdot g^{-1}(2^{\Delta v}) \bigr) \bigr]_0^1\ ,
\end{equation}
\noindent where $g$ is the inverse camera response function (assumed to be an inverse gamma curve $g(x) = x^{2.2}$), $I_{\text{exp}}$ is the re-exposed frame $I$, $\Delta v$ is the change in exposure value, and $[\cdot]^{1}_0$ is an 8-bit rounding operator with clipping in the range $[0,1]$.
To avoid further degradation of the training signal which is SDR in nature, no further exposures are taken from the SDR frames to generate more training samples.


\subsection{Loss Function}
The loss function, $\mathcal{L}$, used for optimizing the model, consists of two terms. The first term, $\mathcal{L}_{\delta}$, is the loss responsible for directly optimizing the residual mapping and the second, $\mathcal{L}_{I}$, is responsible for the overall image mapping consistency: 
\begin{equation}
    \mathcal{L} = 
    \mathcal{L}_{\delta}\left(\hat{\delta}, \delta\right) +
    \mathcal{L}_{I}\left( \hat{I}_{\tilde{b}}, I_{\tilde{b}}\right),
    \label{eq:cyclical:loss}
\end{equation}
\noindent where $\hat{\delta} = \net{}(I_{\tilde{h}} )$ is the residual prediction, ${\hat{I}_{\tilde{b}}} = \hat{\delta} I_{\tilde{b}}$ is the resulting base exposure prediction from the higher exposure frames $ I_{\tilde{h}}$ in the dataset. 

The residual loss, $\mathcal{L}_d$, is the $L_2$ loss because we want to penalize large changes in predicting the multiplicative residuals.
The image space loss $\mathcal{L}_{I}= f(I_x, I_y)$, consists of an $L_1$ distance term and a cosine similarity term that helps enforce color consistency:
\begin{equation}
f(I_x, I_y) = \|I_x - I_y \|_1 + \lambda
\biggl( 
1 - \frac{1}{N} \sum_{j=1}^N\frac{I_x^j \cdot I_y^j}{\|I_x^j\|_2\|I_y^j\|_2}
\biggr)
\end{equation}
\noindent where $N$ is the total number of pixels of the image, $I^j$ is the $j$-th RGB pixel vector of image $I$, and $\lambda$ is a constant factor that adjusts the contribution of the cosine similarity term (in our pilot experiments $\lambda=5$ gave satisfactory results).

%

\subsection{Model}

\begin{figure}[t]
    \centering
    \includegraphics[width=1.0\linewidth]{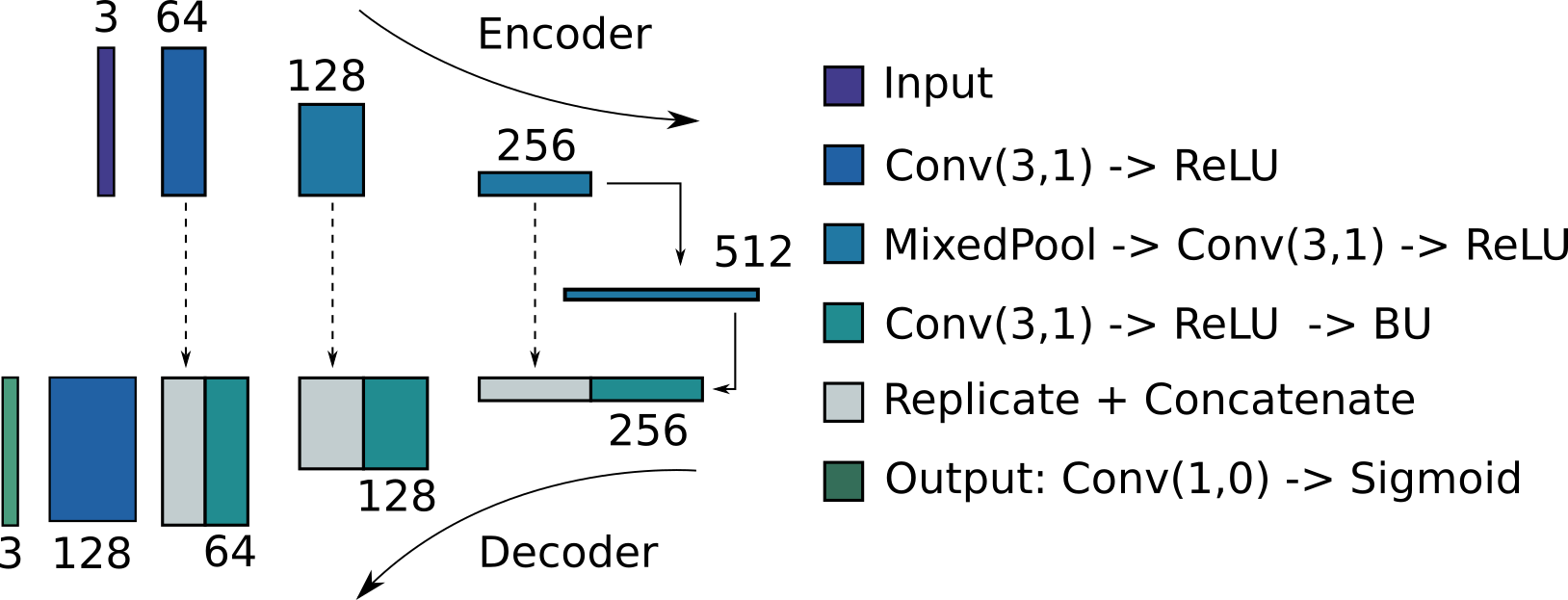}
    \caption{\small{Diagram of the network architecture used by \net. Conv($k$,$p$) is a 2D convolutional layer with kernel size $k$ and padding $p$. BU denotes bilinear upsampling by a factor of 2.}}
    \label{fig:model}
\end{figure}

\net{} is based on the UNet architecture~\cite{Ronneberger+2015} and consists of an encoder and a decoder part with skip-connections and 9 convolutional layers in total, see Figure \ref{fig:model}.
The standard ReLU activation is used but the use of batch normalization (BN) is avoided. This is because the BN layers were found to cause blob-like artifacts in our initial experiments, likely due to the change in input statistics when running at inference mode using a different exposure value as input. Fixed-Pooling~\cite{lee2016generalizing}, which is a learnable combination of max-pooling and average pooling, is used for downsampling in the encoder, while bilinear upsampling is used in the encoder. Figure~\ref{fig:model} shows a diagram of the architecture.
\section{Results}

In this section we present quantitative and qualitative results against fully supervised state of the art methods: Santos et al.\cite{Santos+2020}, Eilertsen et al.\cite{Eilertsen+2017} using retrained parameters for temporal coherency\cite{Eilertsen+2019}, Endo et al.~\cite{Endo+2017}, and Marnerides et al.\cite{Marnerides+2018}. The methods will be referred to as: SAN (Santos et al.), EIL ( Eilerstein et al.), EXP (Marnerides et al.), DRT (Endo et al.), and OUR (the presented method).
We do not compare with any zero-shot, self-supervised or semi-supervised methods, as to the best of our knowledge, none exist for inverse tone mapping. Note that we used the original authors' source code and weights for all these methods. 

For evaluation, we gathered 46 HDR videos from two popular HDR video datasets: the Stuttgart HDR Video dataset (STU) \cite{Froehlich+2014} and the UBC DML-HDR dataset (UBC) \cite{Dehkordi+2014}. It is important to note that frames from these HDR videos were part of the training set of the state-of-the-art methods we compared against; but due to the scarcity of true HDR videos, not many datasets are available and the community will, commonly, use similar datasets. This is largely unavoidable and may have a detrimental effect to our method in the comparisons.
To demonstrate our method in fairer conditions, we employed a set of 4 HDR videos from the IC-1005 project (COST)\footnote{https://www.cost.eu/actions/IC1005/}, which are available by request and to the best of our knowledge have not been used in training of any of the compared state-of-the-art methods.

\subsection{Training: Video Generation}

For our method, the training for each video was performed independently, on a Linux machine (Ubuntu 18.04) equipped with an Intel CPU Core i7-7800X ($3.50$~GHz) with 64~GB of memory and an NVIDIA GeForce 3080 GPU with 10~GB of memory (CUDA 11.3). We implemented our model using the PyTorch 1.9.0 deep-learning framework.

To train our network, we employed mini-batch stochastic gradient descent and the Adam update rule \cite{Kingma+2014} with the learning rate set to 0.001. We left the rest of the parameters set to their default values; i.e., $\beta_1=0.9$, $\beta_2=0.999$, and $\epsilon=1e^{-8}$. 
For each of our trained videos, we set the maximum number of epochs to 128. Typically, we reached a plateau of our loss around epoch 64-100.
We trained using batch size of 1 due to memory constraints when training with frame resolutions of $512\times 512$. All the videos are of resolution $512\times512$ both at training and evaluation time.
%
The duration of training depends on the duration of the input video. A linear relationship exists between the duration of a video and a single training epoch. For example, a four second video requires four seconds to train one epoch on the employed machine.

In terms of evaluation time (the time required for expanding an SDR frame), the model maintains the linear complexity of UNets (i.e., linearly proportional to the number of input pixels). To generate four images at higher and lower f-stops (\textit{i.e.} -4 fstops, -2 fstops, +2 fstops, and +4 fstops) from the input frames at HD resolution (i.e., $1920\times 1080$ the model requires 317 milliseconds of computation.

\begin{table}[ht]
    \centering
    \small{
    \begin{tabular}{|l|C{1.5cm}|C{1.5cm}|C{1.5cm}|}
    \hline
    \multicolumn{4}{|c|}{STU  dataset}\\
    \hline
    \textbf{Method} & \textbf{PU-PSNR} & \textbf{PU-SSIM} & 
    \textbf{HDR-VDP2.2}\\
    \hline
OUR & \bluet{34.9619} &   \redt{0.9851} & 57.8989 \\
SAN & 33.5083  & 0.9258 &  \bluet{60.6266}\\
EIL &  \redt{35.0564}  &  \bluet{0.9267} &  \redt{61.1596} \\
EXP & 26.6346  &  0.8237 & 54.0230   \\    
DRT & 17.4895  & 0.4184 & 44.4383  \\
    \hline
    \multicolumn{4}{|c|}{UBC  dataset}\\
    \hline
OUR & \redt{39.7362}  &  \redt{0.9917} & 59.9136\\
SAN & 33.2781  &  \bluet{0.9845} & \redt{64.5952} \\
EIL & \bluet{33.3537}  & 0.9831 & \bluet{63.7712} \\
EXP & 22.7719 & 0.8627 & 56.8770\\    
DRT & 17.7969  & 0.7362 & 50.2718\\
    \hline
    \multicolumn{4}{|c|}{COST  dataset}\\
    \hline
OUR & \redt{45.1735}  &  \redt{0.9956} &  \redt{70.4994}\\
SAN & \bluet{31.5296} &  \bluet{0.9725}  &  68.3405\\
EIL & 31.4386  &  0.9721 & \bluet{68.8438}\\
EXP & 28.9225  & 0.9399 & 62.4470\\    
DRT & 14.7202   &   0.7477 & 51.1164\\
    \hline
    \end{tabular}
    }
    \caption{\small{This table reports the PU-PSNR, PU-SSIM, and HDR-VDP2.2 (higher values are better for all metrics),  mean values. The red font color is for the best method, and the blue one is for the second-best one.}}
    \label{tab:results:dr}
\end{table}

\subsection{Quantitative}
\label{sec:results:quantitative}
For quantitative results, the generated inverse tone mapped videos for all the methods (including ours) were compared with the ground truth using standard metrics for HDR applications and inverse tone mapping: HDR-VDP2.2\cite{Narwaria+2015}, PU-PSNR\cite{Aydin+2008}, and PU-SSIM\cite{Aydin+2008}; for all these metrics the higher values correspond to better performance. PU-PSNR and PU-SSIM are modified versions of PSNR and SSIM\cite{Wang+2003} where input images are PU-encoded\cite{Mantiuk+21}, before being processed by the metric, to handle how the human visual system perceives HDR data. HDR values follow the VESA DisplayHDR1400 standard\footnote{https://displayhdr.org/} that has a peak luminance of $1,400$~cd/m$^2$ and a black level of $0.02$~cd/m$^2$.

To generate, SDR input frames, we computed a temporally stable (by exponential smoothing) automatic exposure (i.e., mean luminance of the frame) at each HDR frame as:
\begin{equation}
    I^i_{b} = \biggl[ \bigl(I^i_\text{HDR} \cdot 2^{f^i}\bigr)^\frac{1}{2.2}  \biggr]_0^{1}
\end{equation}
\noindent where $I^i_\text{HDR}$ is the $i$-th HDR frame, $I^i_b$ is the SDR frame, $f^i$ is the exposure value (in f-stops) for the $i$-th HDR frame, and $[\cdot]^{1}_0$ is the same rounding and clipping operator as in Equation~\ref{eqn:exposure}.

%

Table~\ref{tab:results:dr} summarizes the comparisons for PU-PSNR, PU-SSIM and HDR-VDP2.2. Means are computed across all videos for each method and metric. 
These results show that our method works well in terms of PU-PSNR and PU-SSIM against the state-of-the-art for the STU and UBC datasets. Although it has reasonable results for HDR-VDP2.2, our method does not outperform SAN and EIL. 
These results reflect the same ranking as seen in Santos et al.'s work \cite{Santos+2020}.
It is important to note that, apart from our proposed method, the other methods were trained using the STU and UBC datasets, which explains their performance with this metric. 

However, when comparing our method against the state-of-the-art using a dataset that was not used by the other methods during their training, \textit{i.e.} COST in this case, our method performs significantly better than the state-of-the-art across all metrics. 
This shows the applicability of the proposed method to generalize well as can be seen when comparing results across unseen datasets. 

\subsection{Visual Inspection}
We also show qualitative results, comparing our method with the state of the art and the original HDR frames. For all methods, the input is a 0 f-stop image from the HDR ground truth (GT). We refer the reader to the additional video for visual comparisons of videos at different exposures.

\begingroup
\setlength\tabcolsep{0pt}
\renewcommand{\arraystretch}{0}
\begin{figure}
    \centering
    \begin{tabular}{cccccc}
    &\textbf{-4 fstop} & \textbf{-2 fstop} & \textbf{0 fstop} & \textbf{+2 fstop}& \textbf{+4 fstop}\\
    \raisebox{1.5\height}{\rotatebox[origin=c]{90}{OUR}\ }&
    \includegraphics[width=0.18\linewidth]{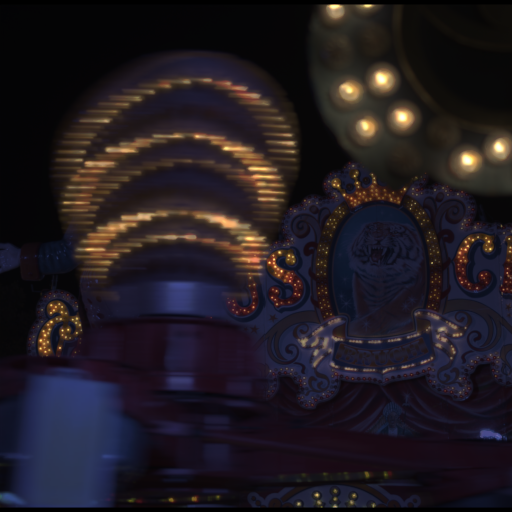}&
    \includegraphics[width=0.18\linewidth]{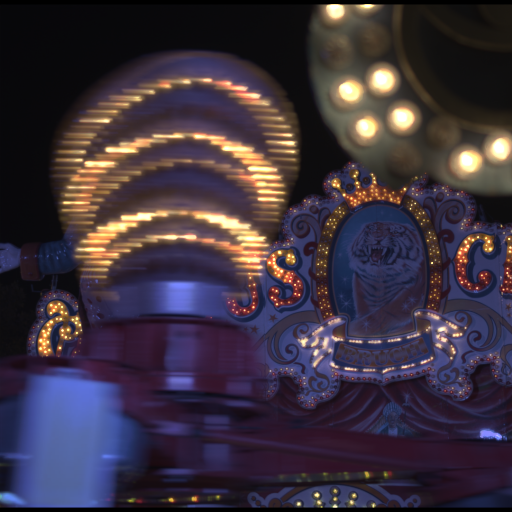}&
    \includegraphics[width=0.18\linewidth]{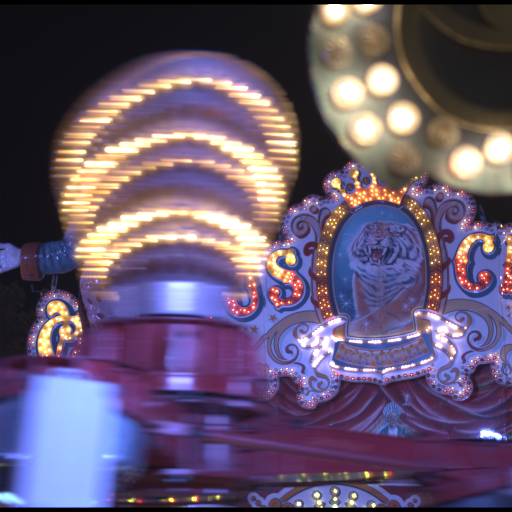}&
    \includegraphics[width=0.18\linewidth]{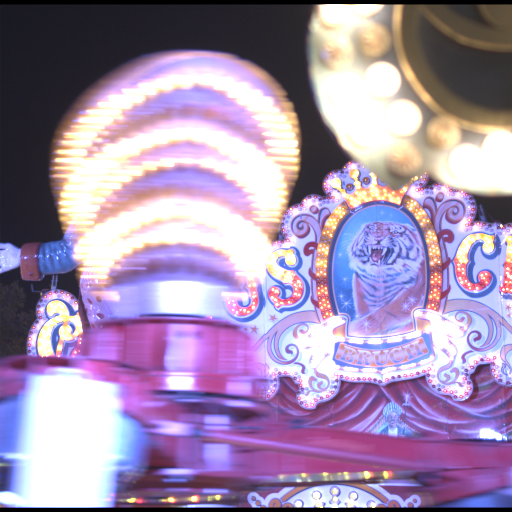}&
    \includegraphics[width=0.18\linewidth]{img/carousel2/carousel_fireworks_02_000958_our_+2.png}\\
    \raisebox{1.5\height}{\rotatebox[origin=c]{90}{SAN
    }}&
    \includegraphics[width=0.18\linewidth]{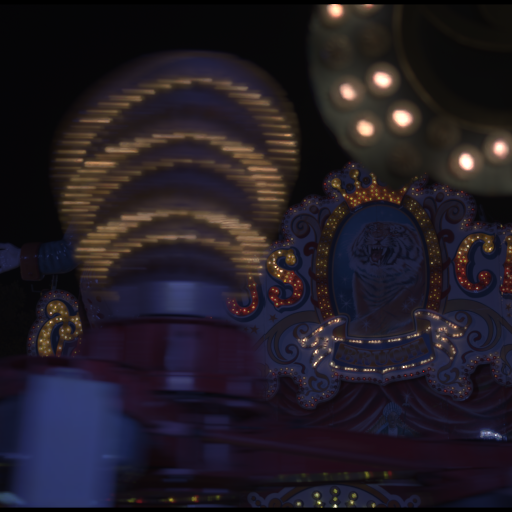}&
    \includegraphics[width=0.18\linewidth]{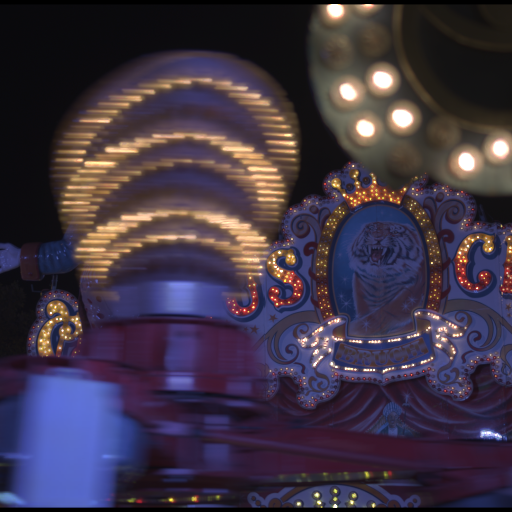}&
    \includegraphics[width=0.18\linewidth]{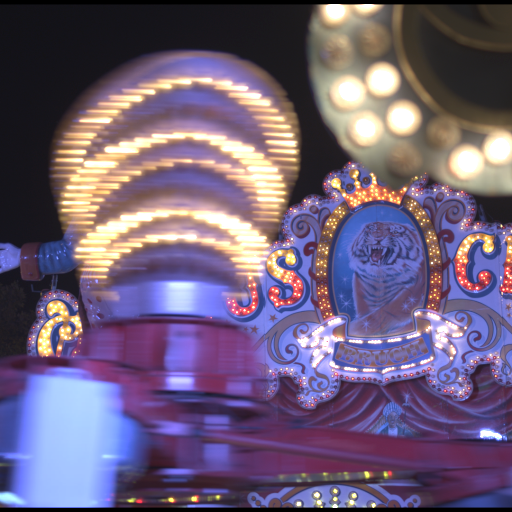}&
    \includegraphics[width=0.18\linewidth]{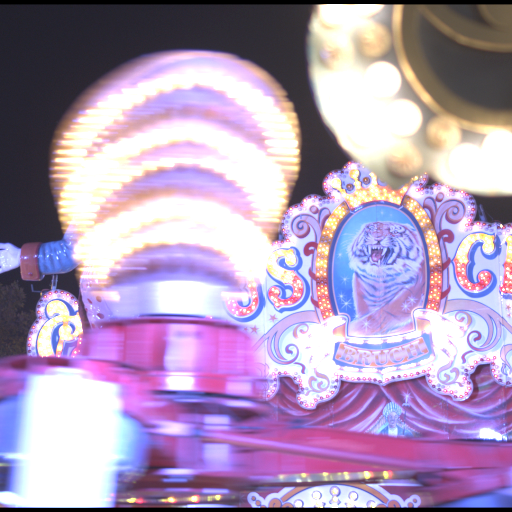}&
    \includegraphics[width=0.18\linewidth]{img/carousel2/carousel_fireworks_02_000958_s_+2.png}\\
    \raisebox{1.5\height}{\rotatebox[origin=c]{90}{EIL
    }}&
    \includegraphics[width=0.18\linewidth]{img/carousel2/carousel_fireworks_02_000958_s_-4.png}&
    \includegraphics[width=0.18\linewidth]{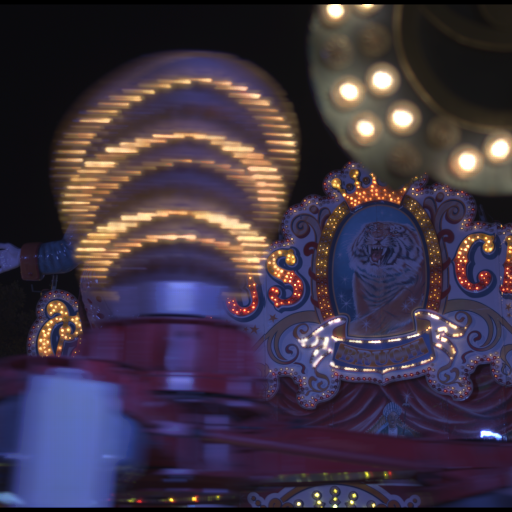}&
    \includegraphics[width=0.18\linewidth]{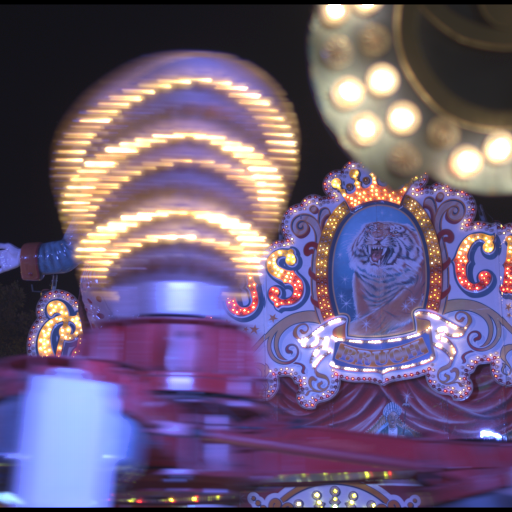}&
    \includegraphics[width=0.18\linewidth]{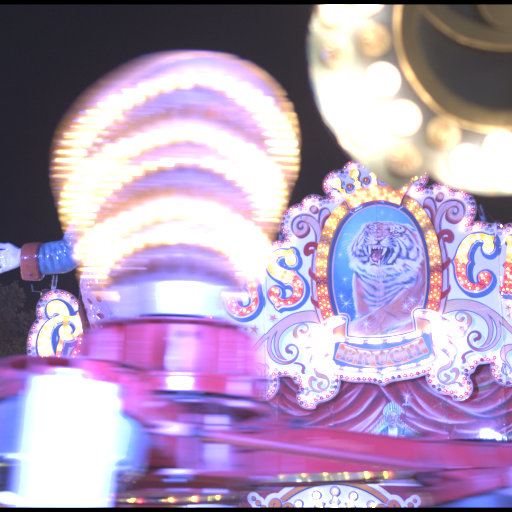}&
    \includegraphics[width=0.18\linewidth]{img/carousel2/carousel_fireworks_02_000958_e_+2.png}\\
    \raisebox{1.5\height}{\rotatebox[origin=c]{90}{EXP
    }}&
    \includegraphics[width=0.18\linewidth]{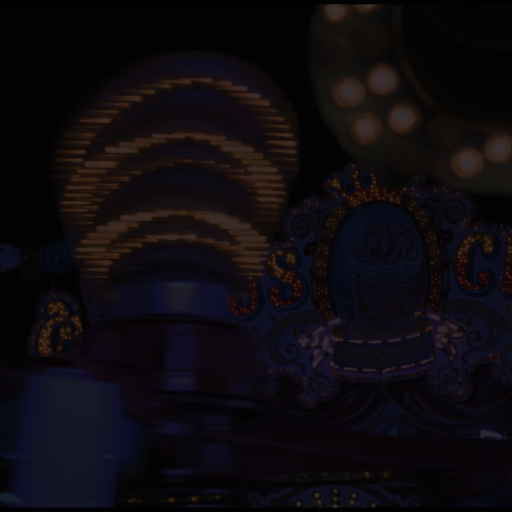}&
    \includegraphics[width=0.18\linewidth]{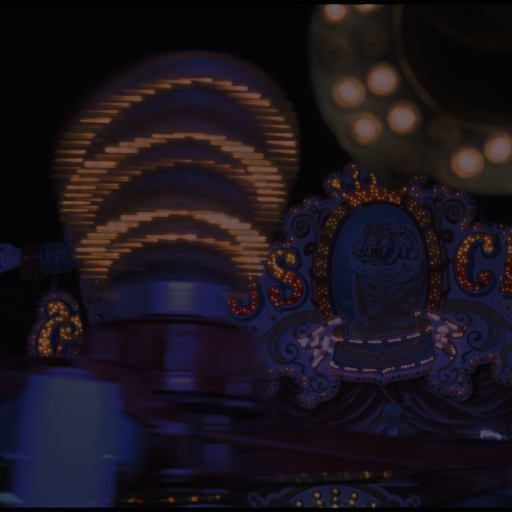}&
    \includegraphics[width=0.18\linewidth]{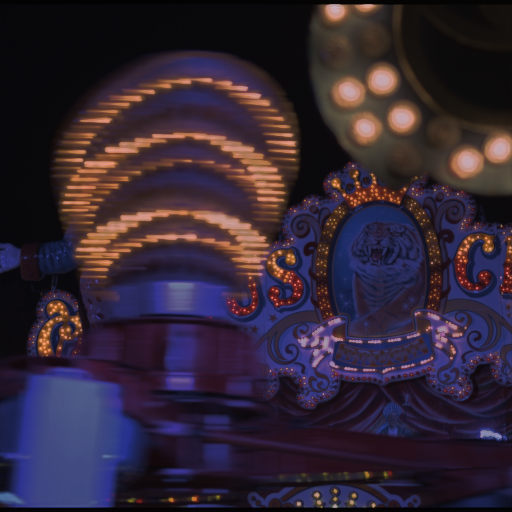}&
    \includegraphics[width=0.18\linewidth]{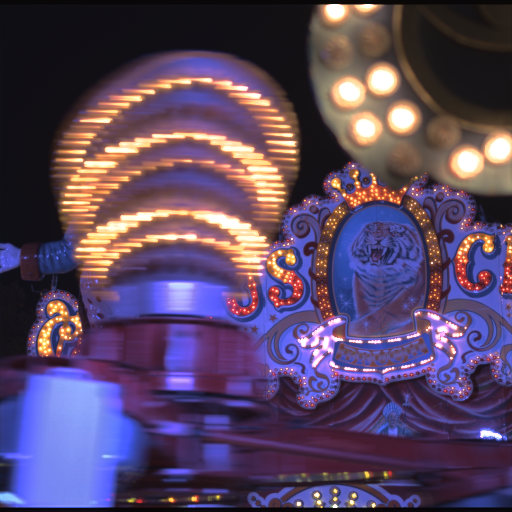}&
    \includegraphics[width=0.18\linewidth]{img/carousel2/carousel_fireworks_02_000958_en_+2.png}\\
    \raisebox{1.5\height}{\rotatebox[origin=c]{90}{DRT
    }}&
    \includegraphics[width=0.18\linewidth]{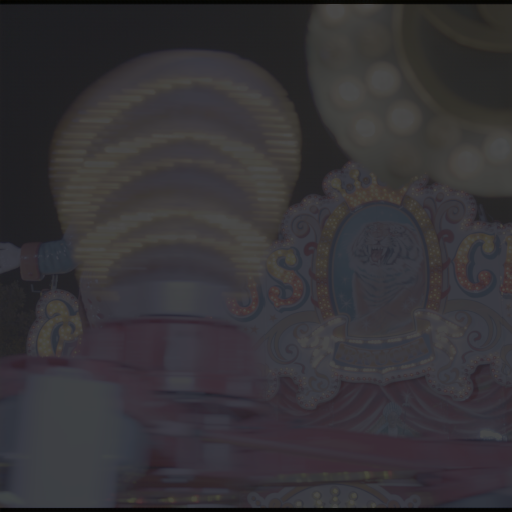}&
    \includegraphics[width=0.18\linewidth]{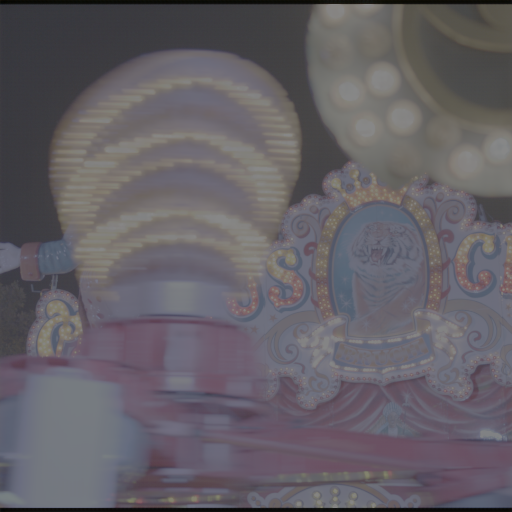}&
    \includegraphics[width=0.18\linewidth]{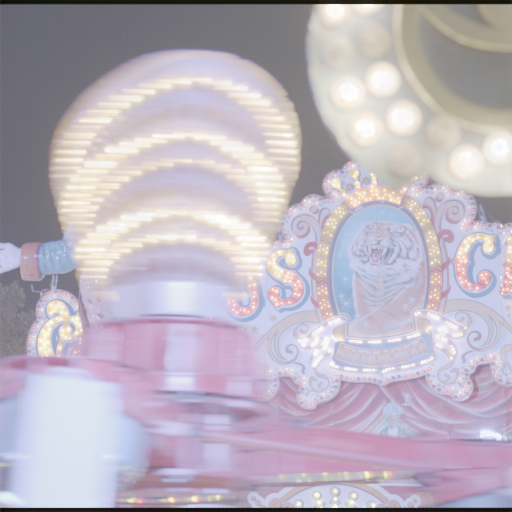}&
    \includegraphics[width=0.18\linewidth]{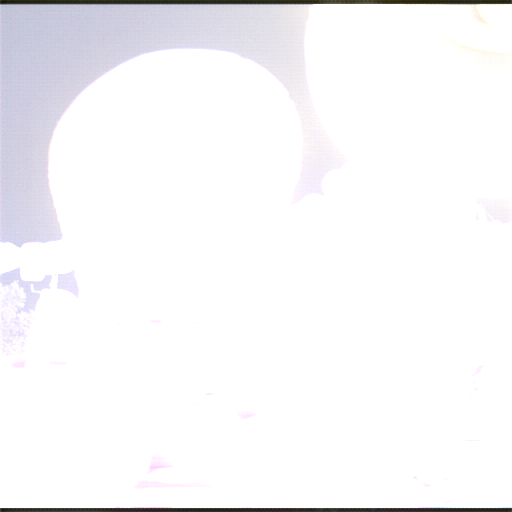}&
    \includegraphics[width=0.18\linewidth]{img/carousel2/carousel_fireworks_02_000958_dr_+2.png}\\
    \raisebox{2\height}{\rotatebox[origin=c]{90}{GT
    }}&
    \includegraphics[width=0.18\linewidth]{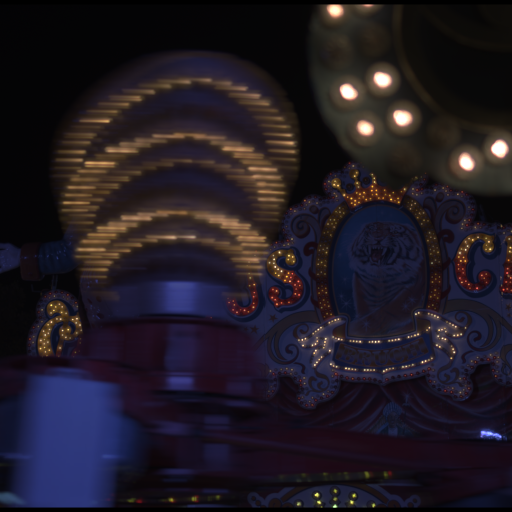}&
    \includegraphics[width=0.18\linewidth]{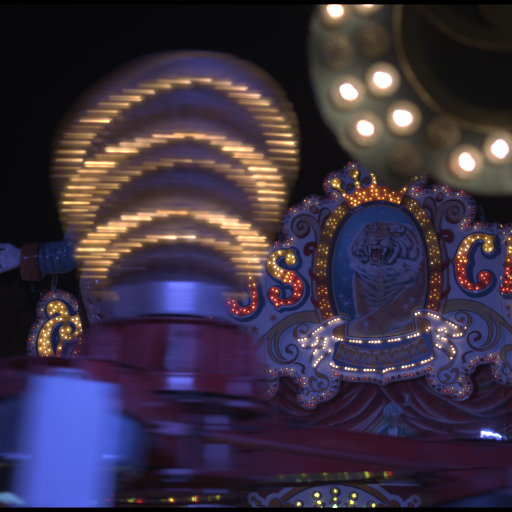}&
    \includegraphics[width=0.18\linewidth]{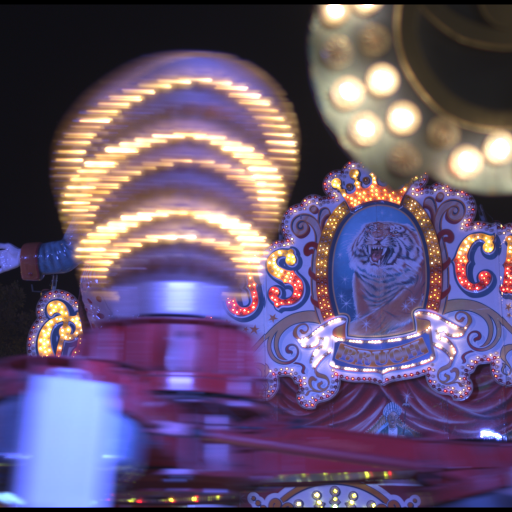}&
    \includegraphics[width=0.18\linewidth]{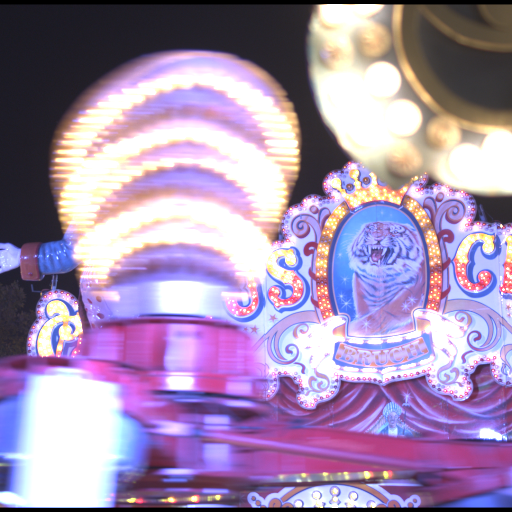}&
    \includegraphics[width=0.18\linewidth]{img/carousel2/carousel_fireworks_02_000958_+2.png}
    \end{tabular}    
    \caption{\small{A visual comparison of all tested methods. The frame is part of the sequence Carousel\_Fireworks\_02 \cite{Froehlich+2014}. This shows that OUR method can reconstruct details in the light sources to other methods yet only relies on the original SDR content.}}
    \label{fig:visual:carousel02}
\end{figure}
\endgroup

Figure \ref{fig:teaser} and Figure \ref{fig:visual:carousel02} show an example of our method applied to a challenging scene showing our method reconstructs detail in overexposed areas of the frames including reconstructing texture and colors even in the presence of motion blur.

Figure~\ref{fig:visual:fireplace} is a challenging example where there is rapid motion and texture details, colors, and a significant lack of dynamic range in the input. Our method can generate similar details in terms of color, dynamic range, and texture. When compared to EIL, for example, our method manages to recover more texture and details in the flames, obtaining similar results to SAN.

\begingroup
\setlength\tabcolsep{0pt}
\renewcommand{\arraystretch}{0}
\begin{figure}
    \centering
    \begin{tabular}{cccccc}
    &\textbf{-4 fstop} & \textbf{-2 fstop} & \textbf{0 fstop} & \textbf{+2 fstop}& \textbf{+4 fstop}\\
    \raisebox{1.5\height}{\rotatebox[origin=c]{90}{OUR}\ }&
    \includegraphics[width=0.18\linewidth]{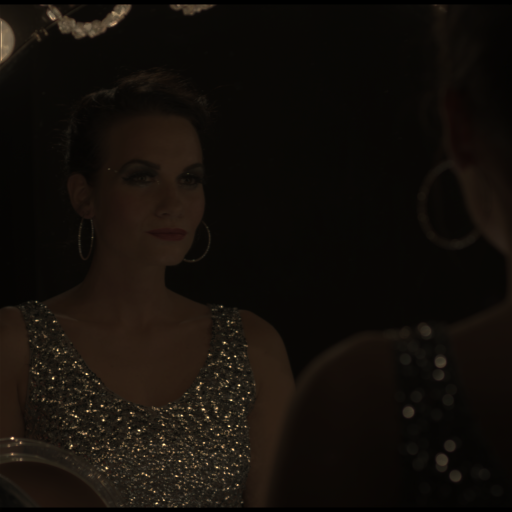}&
    \includegraphics[width=0.18\linewidth]{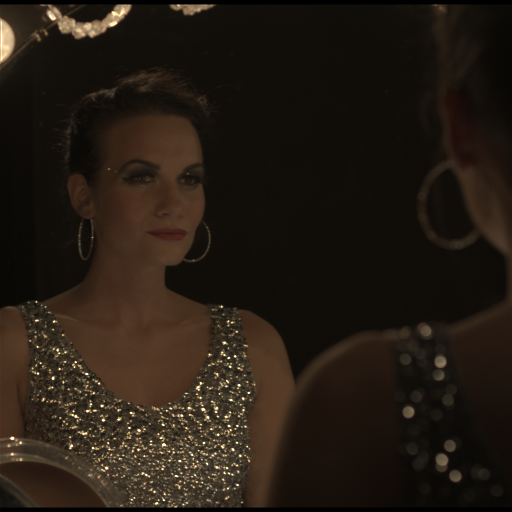}&
    \includegraphics[width=0.18\linewidth]{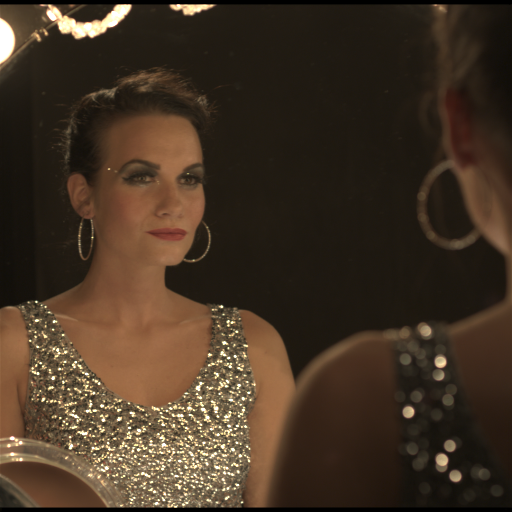}&
    \includegraphics[width=0.18\linewidth]{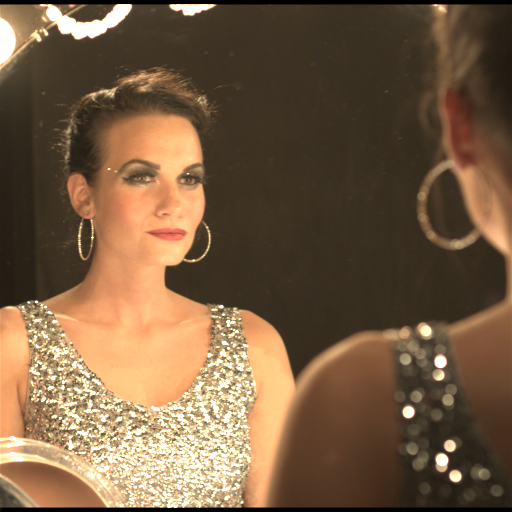}&
    \includegraphics[width=0.18\linewidth]{img/visual_show_girl_01/showgirl_01_000695_our_+2.png}\\
    \raisebox{1.5\height}{\rotatebox[origin=c]{90}{SAN
    }}&
    \includegraphics[width=0.18\linewidth]{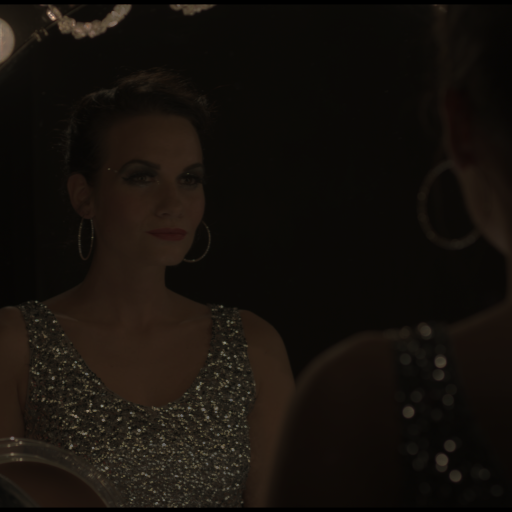}&
    \includegraphics[width=0.18\linewidth]{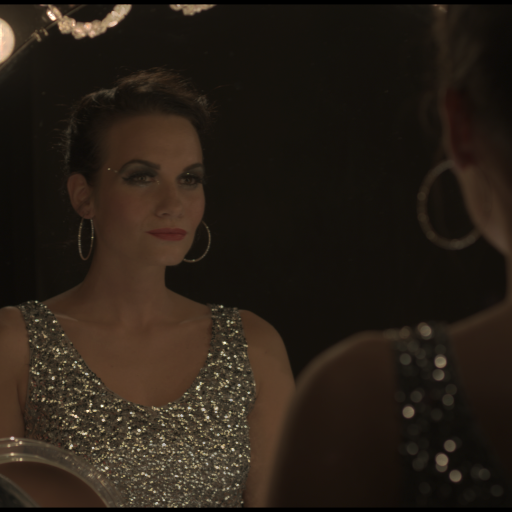}&
    \includegraphics[width=0.18\linewidth]{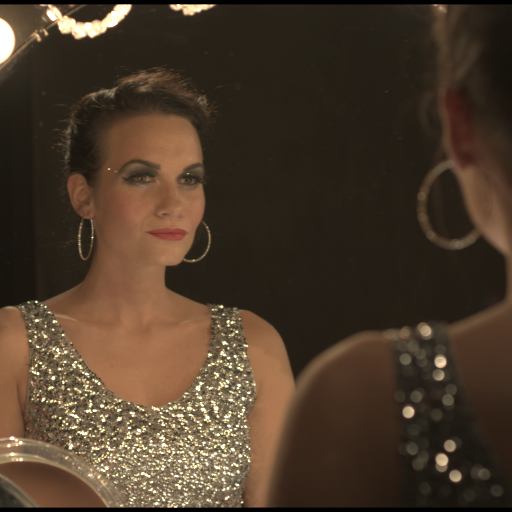}&
    \includegraphics[width=0.18\linewidth]{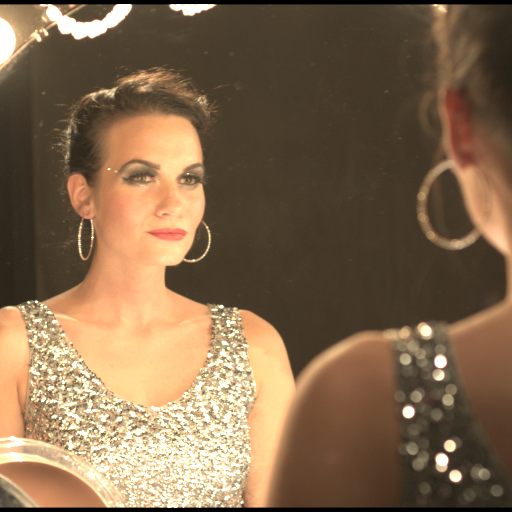}&
    \includegraphics[width=0.18\linewidth]{img/visual_show_girl_01/showgirl_01_000695_s_+2.png}\\
    \raisebox{1.5\height}{\rotatebox[origin=c]{90}{EIL
    }}&
    \includegraphics[width=0.18\linewidth]{img/visual_show_girl_01/showgirl_01_000695_s_-4.png}&
    \includegraphics[width=0.18\linewidth]{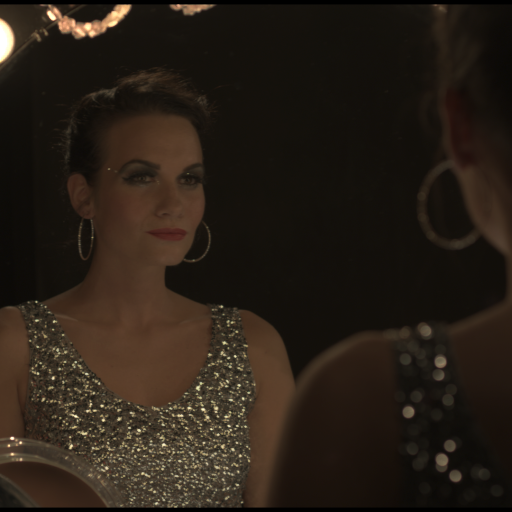}&
    \includegraphics[width=0.18\linewidth]{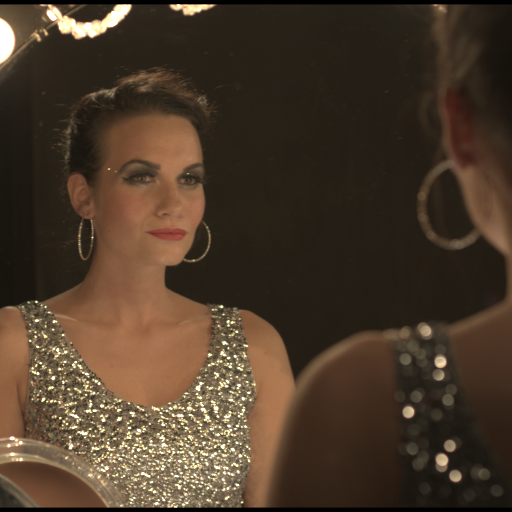}&
    \includegraphics[width=0.18\linewidth]{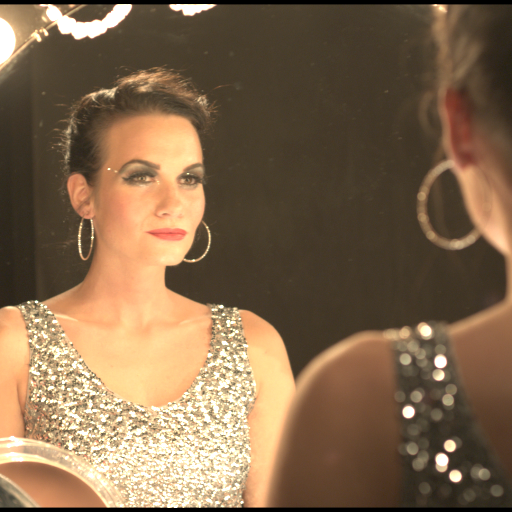}&
    \includegraphics[width=0.18\linewidth]{img/visual_show_girl_01/showgirl_01_000695_e_+2.png}\\
    \raisebox{1.5\height}{\rotatebox[origin=c]{90}{EXP
    }}&
    \includegraphics[width=0.18\linewidth]{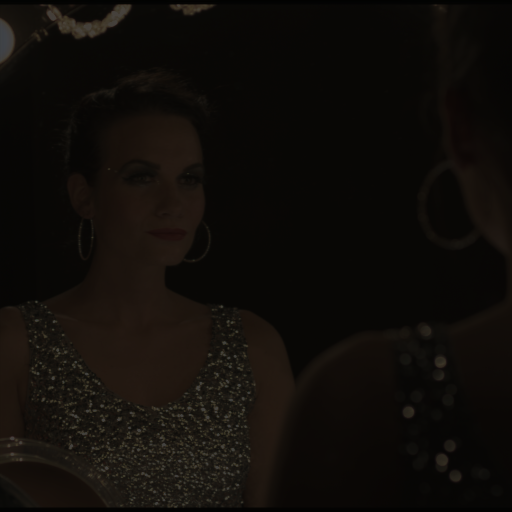}&
    \includegraphics[width=0.18\linewidth]{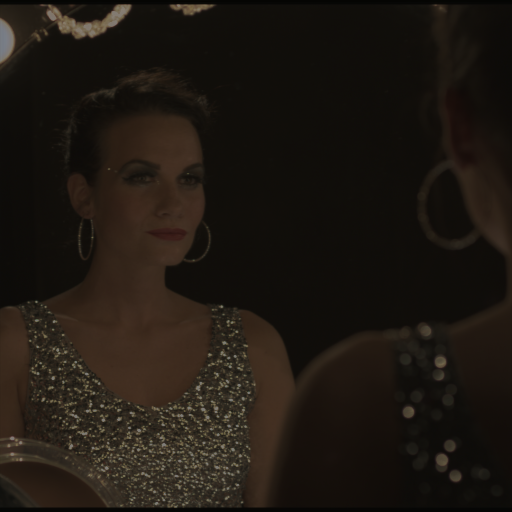}&
    \includegraphics[width=0.18\linewidth]{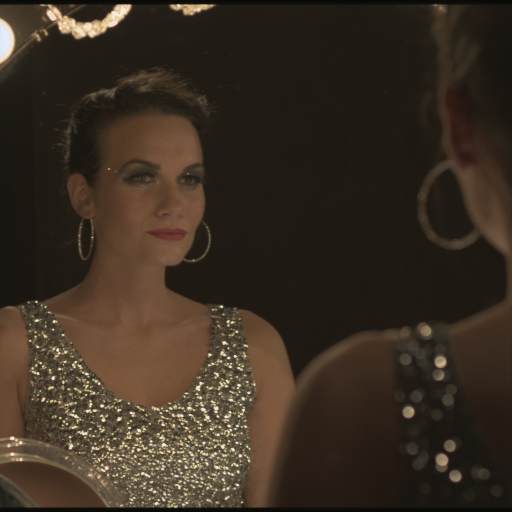}&
    \includegraphics[width=0.18\linewidth]{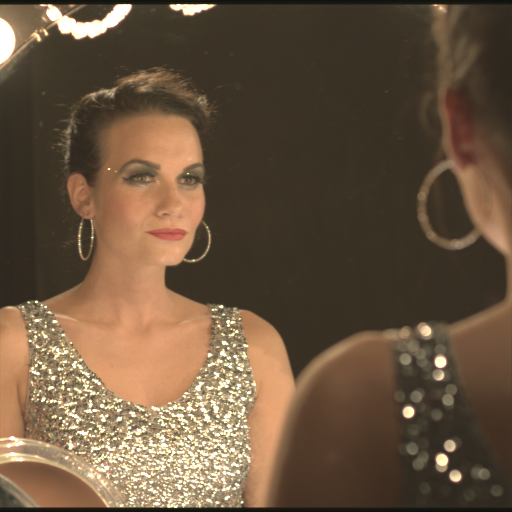}&
    \includegraphics[width=0.18\linewidth]{img/visual_show_girl_01/showgirl_01_000695_en_+2.png}\\
    \raisebox{1.5\height}{\rotatebox[origin=c]{90}{DRT
    }}&
    \includegraphics[width=0.18\linewidth]{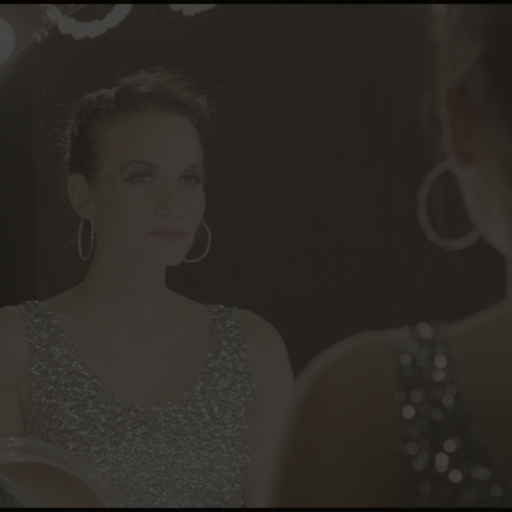}&
    \includegraphics[width=0.18\linewidth]{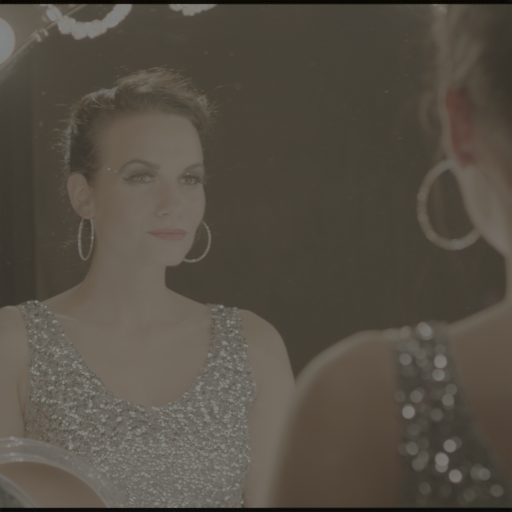}&
    \includegraphics[width=0.18\linewidth]{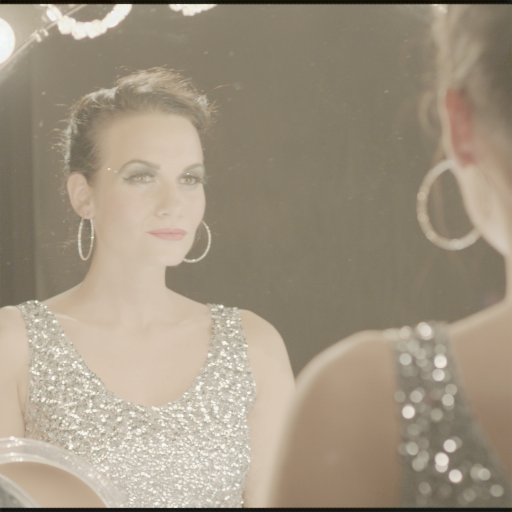}&
    \includegraphics[width=0.18\linewidth]{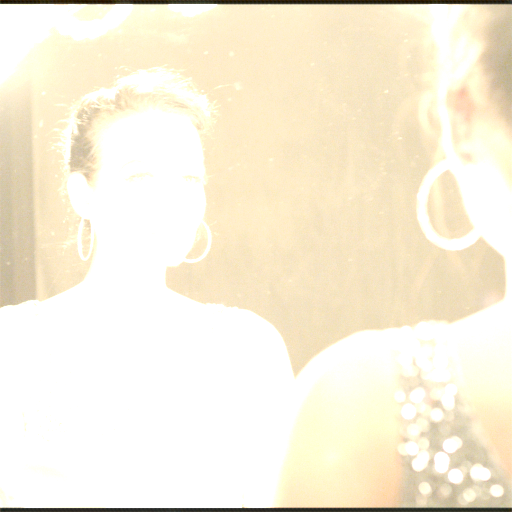}&
    \includegraphics[width=0.18\linewidth]{img/visual_show_girl_01/showgirl_01_000695_dr_+2.png}\\
    \raisebox{2\height}{\rotatebox[origin=c]{90}{GT
    }}&
    \includegraphics[width=0.18\linewidth]{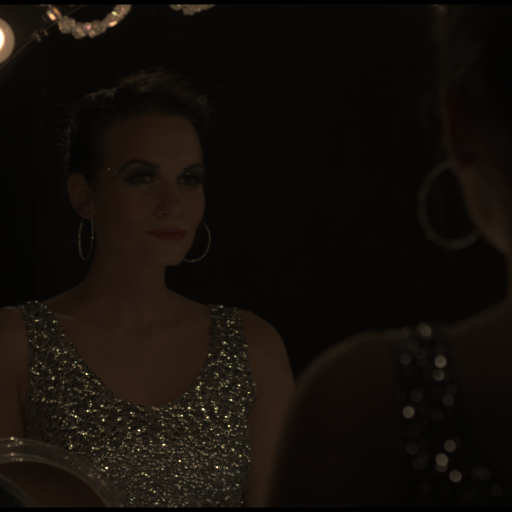}&
    \includegraphics[width=0.18\linewidth]{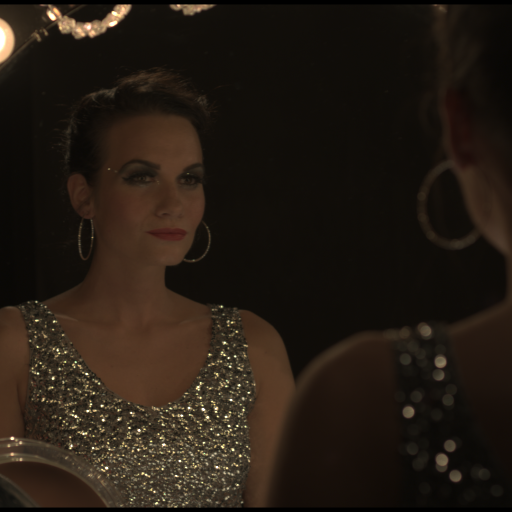}&
    \includegraphics[width=0.18\linewidth]{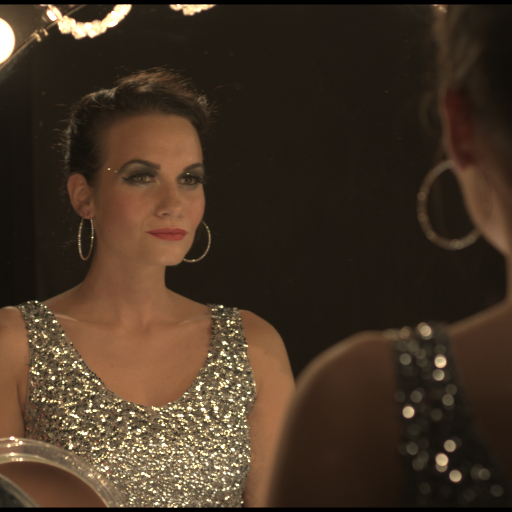}&
    \includegraphics[width=0.18\linewidth]{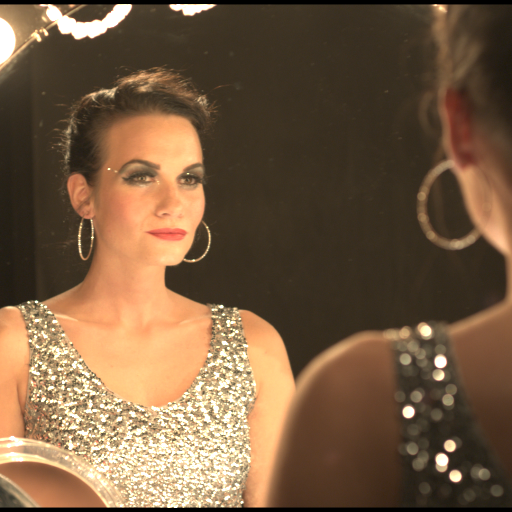}&
    \includegraphics[width=0.18\linewidth]{img/visual_show_girl_01/showgirl_01_000695_+2.png}
    \end{tabular}    
    \caption{\small{A visual comparison of all tested methods. The frame is part of the sequence Showgirl\_01 \cite{Froehlich+2014}. This shows that OUR method can reconstruct details in the lights and dress and performs similarly to other methods yet only relies on the original SDR content.}}
    \label{fig:visual:showgirl01}
\end{figure}
\endgroup

\begin{figure}[ht]
    \centering
    \includegraphics[width=0.3\linewidth]{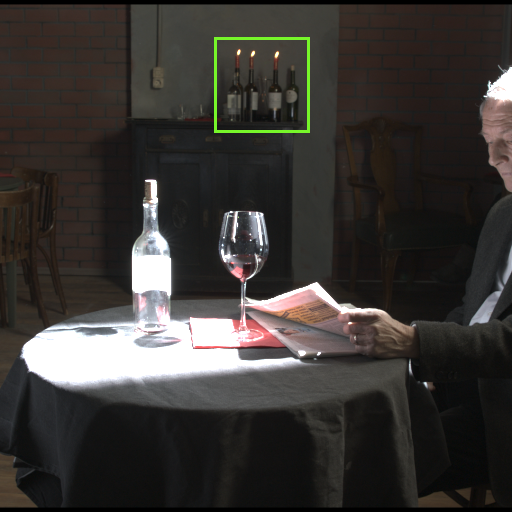}
    \hspace{1pt}
    \includegraphics[width=0.3\linewidth]{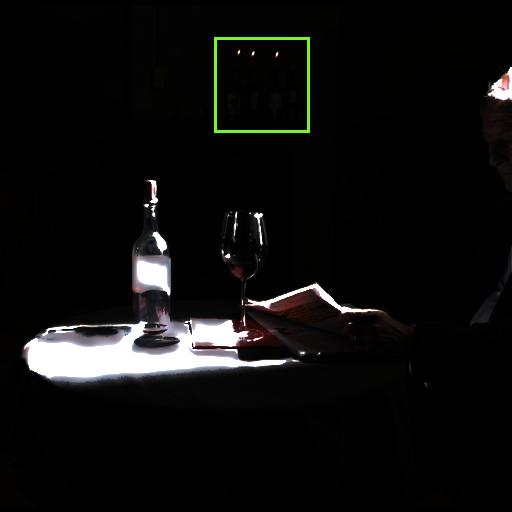}
    \hspace{1pt}
    \includegraphics[width=0.3\linewidth]{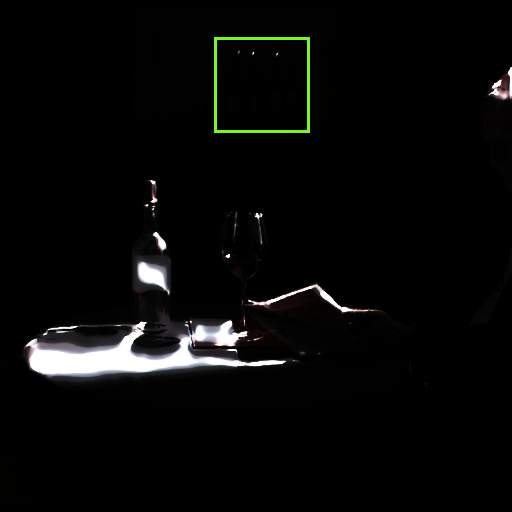}\\
    \includegraphics[width=0.3\linewidth]{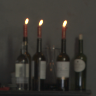}
    \hspace{1pt}
    \includegraphics[width=0.3\linewidth]{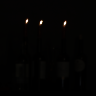}
    \hspace{1pt}
    \includegraphics[width=0.3\linewidth]{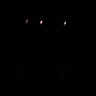}
    \caption{\small{A limitation example (Bistro\_01) \cite{Froehlich+2014}). Here our method can only reconstruct the candles in the green square because they move while the rest of the overexposed scene (the table and the bottles) does not have motion: (a) The input SDR frame. (b) The recovered frame at -2 fstop using our method. (c) The recovered frame at -4 fstop using our method. (d), (e) and (f) show zoomed-in regions of the corresponding green squares in (a), (b), and (c) respectively.
    }}
    \label{fig:limitation}
\end{figure}

Figure \ref{fig:visual:beerfest} has complex light sources that are mostly clipped. Our method can achieve a plausible reconstruction similarly to SAN and EIL. Likewise, Figure \ref{fig:visual:showgirl01} has clipped light sources and texture details on the dress. These are reconstructed well with our method, and the result is comparable to the other state-of-the-art methods.

\begingroup
\setlength\tabcolsep{0pt}
\renewcommand{\arraystretch}{0}
\begin{figure}
    \centering
    \begin{tabular}{cccccc}
    &\textbf{-4 fstop} & \textbf{-2 fstop} & \textbf{0 fstop} & \textbf{+2 fstop}& \textbf{+4 fstop}\\
    \raisebox{1.5\height}{\rotatebox[origin=c]{90}{OUR}\ }&
    \includegraphics[width=0.19\linewidth]{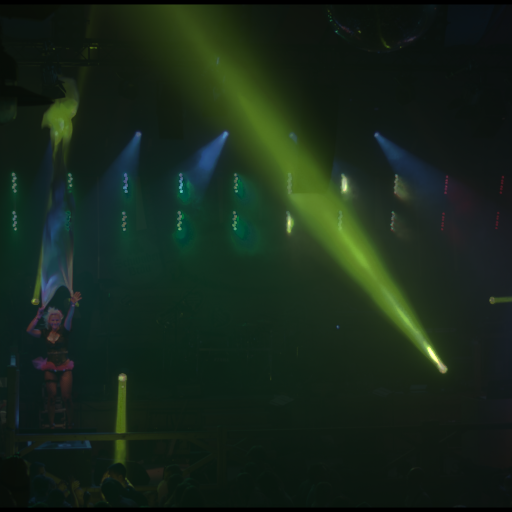}&
    \includegraphics[width=0.19\linewidth]{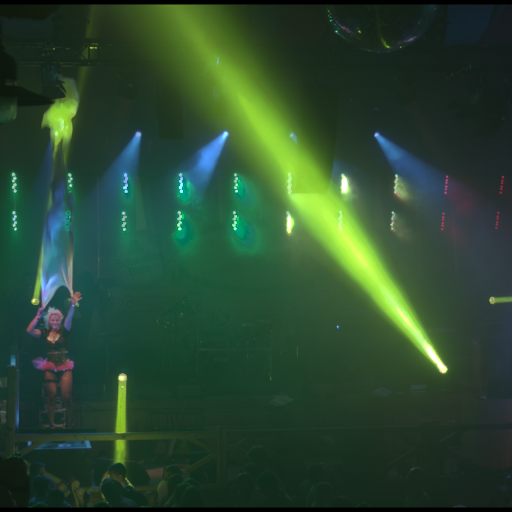}&
    \includegraphics[width=0.19\linewidth]{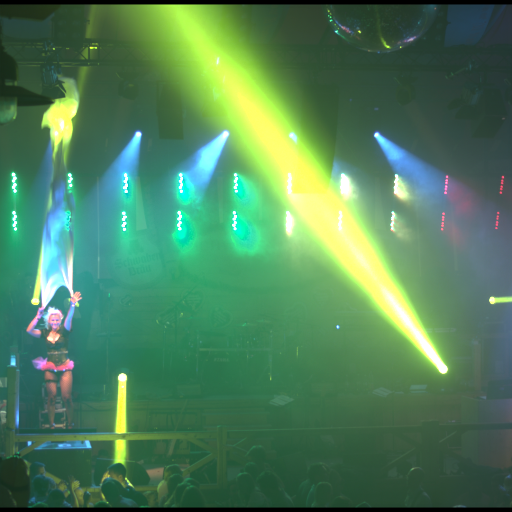}&
    \includegraphics[width=0.19\linewidth]{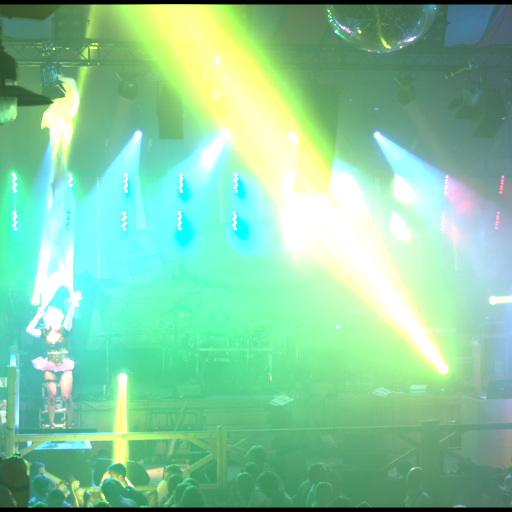}&
    \includegraphics[width=0.19\linewidth]{img/visual_beer/beerfest_lightshow_04_004202_our_+2.png}\\
    \raisebox{1.5\height}{\rotatebox[origin=c]{90}{SAN
    }}&
    \includegraphics[width=0.19\linewidth]{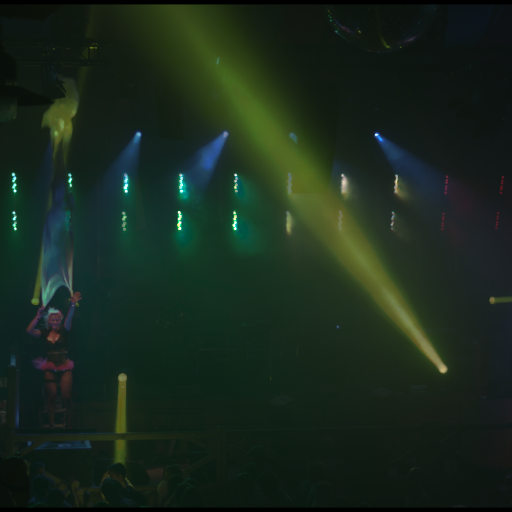}&
    \includegraphics[width=0.19\linewidth]{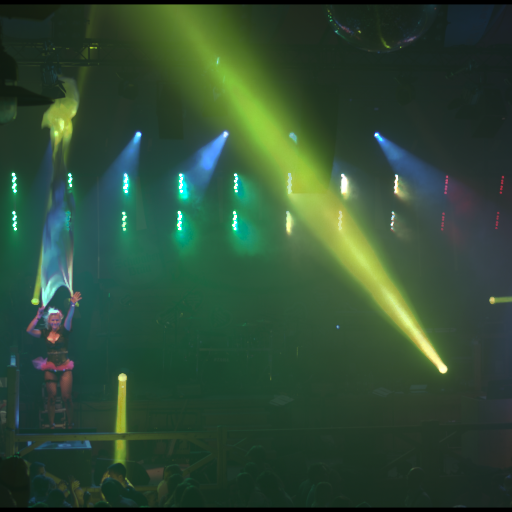}&
    \includegraphics[width=0.19\linewidth]{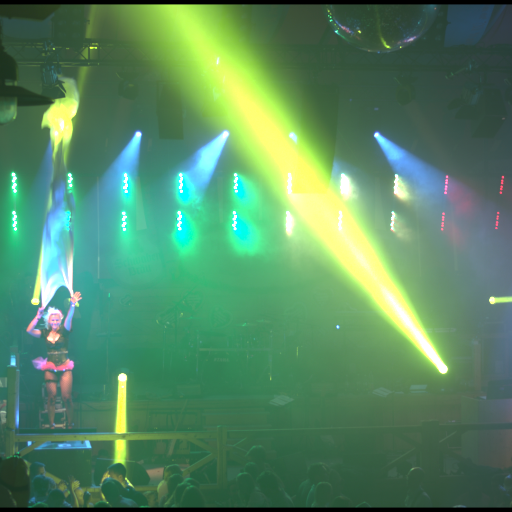}&
    \includegraphics[width=0.19\linewidth]{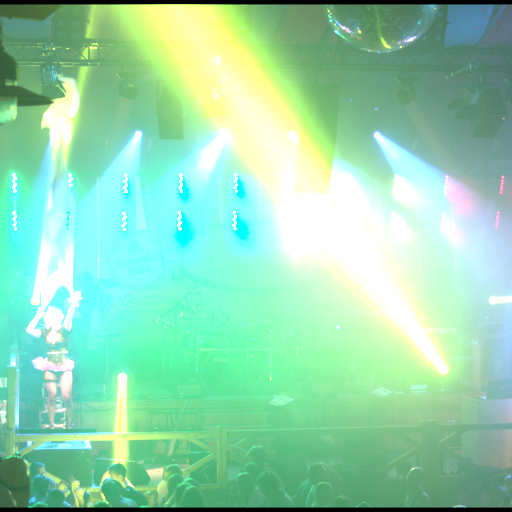}&
    \includegraphics[width=0.19\linewidth]{img/visual_beer/beerfest_lightshow_04_004202_s_+2.png}\\
    \raisebox{1.5\height}{\rotatebox[origin=c]{90}{EIL
    }}&
    \includegraphics[width=0.19\linewidth]{img/visual_beer/beerfest_lightshow_04_004202_s_-4.png}&
    \includegraphics[width=0.19\linewidth]{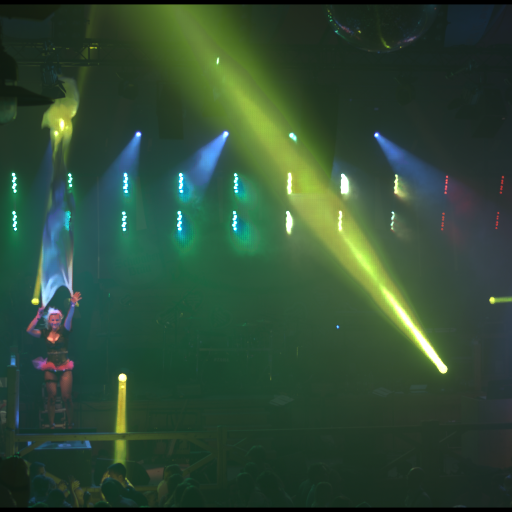}&
    \includegraphics[width=0.19\linewidth]{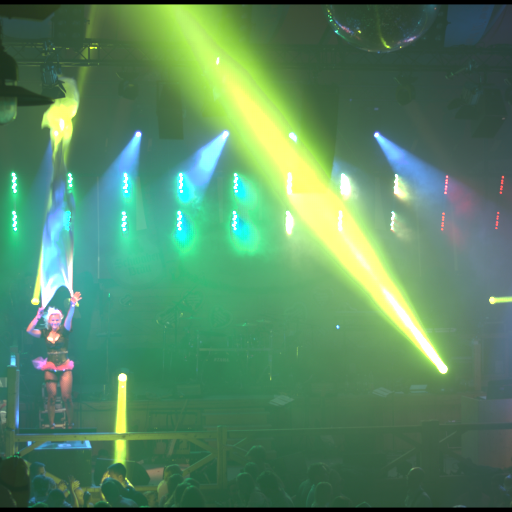}&
    \includegraphics[width=0.19\linewidth]{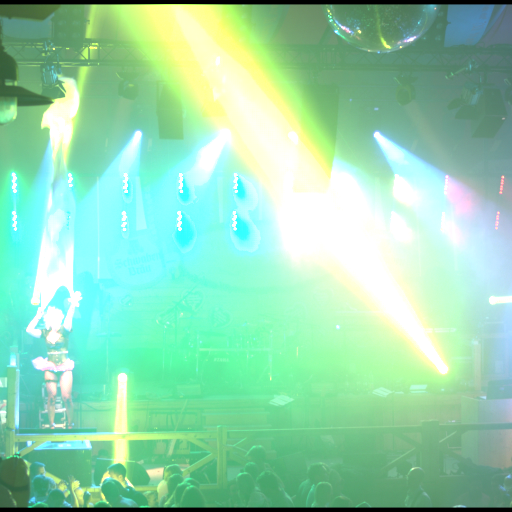}&
    \includegraphics[width=0.19\linewidth]{img/visual_beer/beerfest_lightshow_04_004202_e_+2.png}\\
    \raisebox{1.5\height}{\rotatebox[origin=c]{90}{EXP
    }}&
    \includegraphics[width=0.19\linewidth]{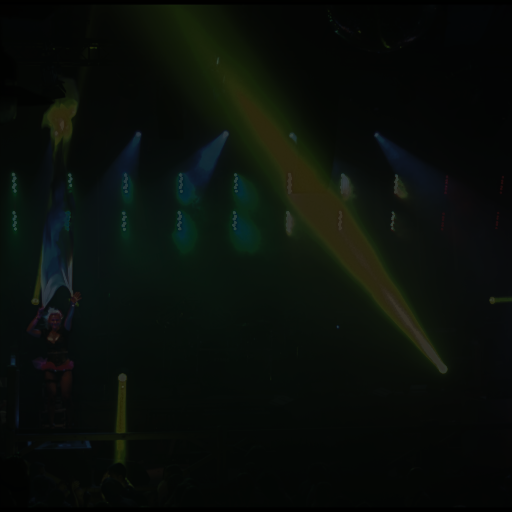}&
    \includegraphics[width=0.19\linewidth]{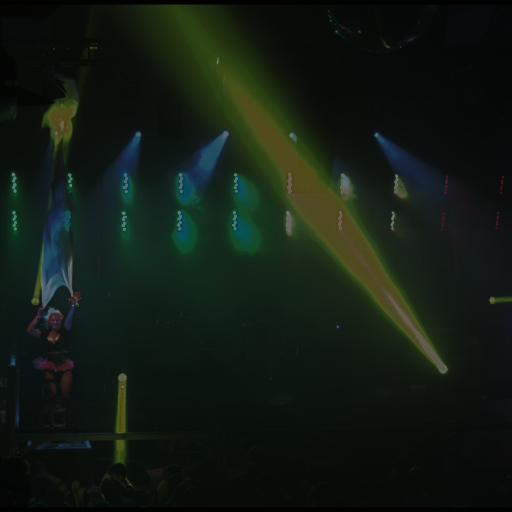}&
    \includegraphics[width=0.19\linewidth]{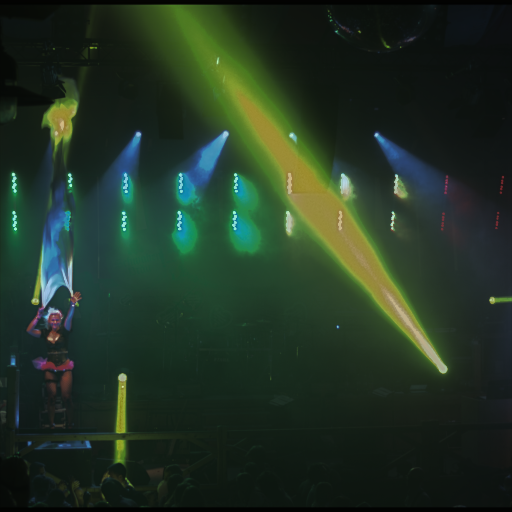}&
    \includegraphics[width=0.19\linewidth]{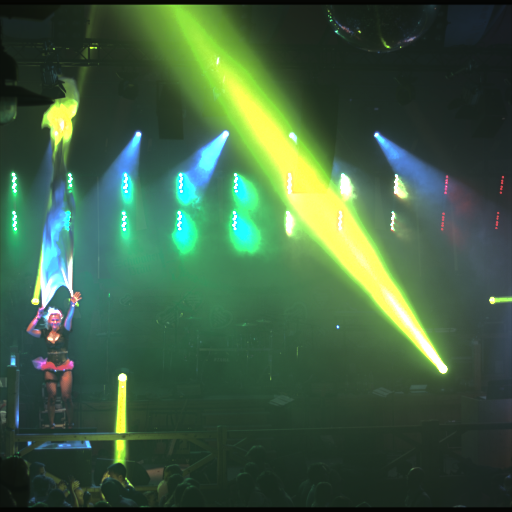}&
    \includegraphics[width=0.19\linewidth]{img/visual_beer/beerfest_lightshow_04_004202_en_+2.png}\\
    \raisebox{1.5\height}{\rotatebox[origin=c]{90}{DRT
    }}&
    \includegraphics[width=0.19\linewidth]{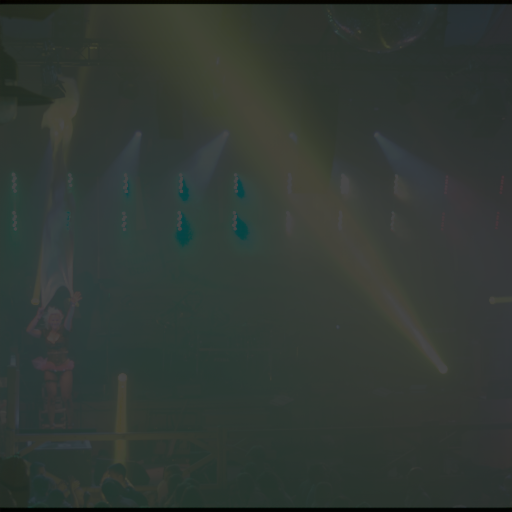}&
    \includegraphics[width=0.19\linewidth]{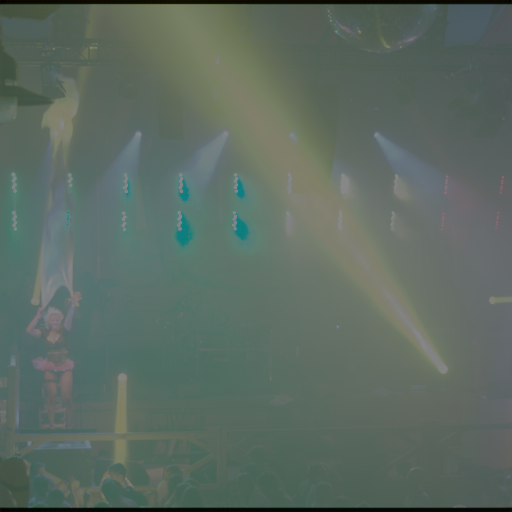}&
    \includegraphics[width=0.19\linewidth]{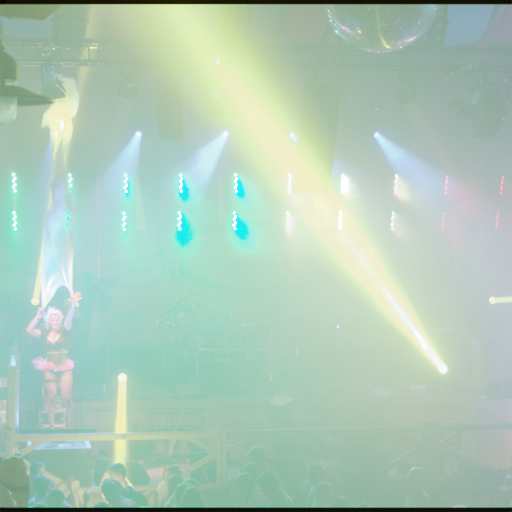}&
    \includegraphics[width=0.19\linewidth]{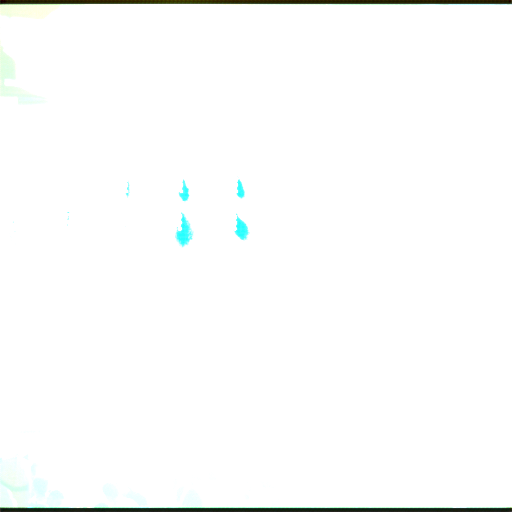}&
    \includegraphics[width=0.19\linewidth]{img/visual_beer/beerfest_lightshow_04_004202_dr_+2.png}\\
    \raisebox{2\height}{\rotatebox[origin=c]{90}{GT
    }}&
    \includegraphics[width=0.19\linewidth]{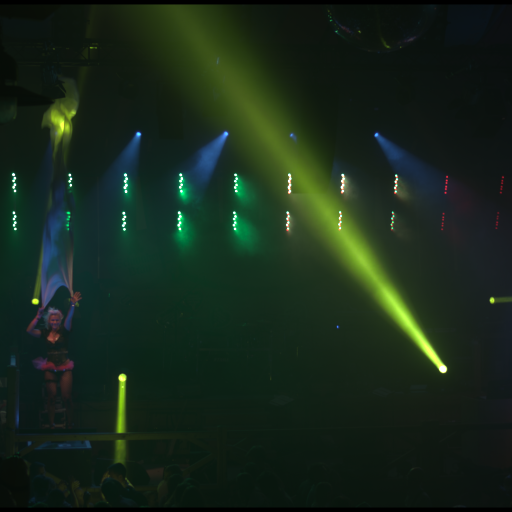}&
    \includegraphics[width=0.19\linewidth]{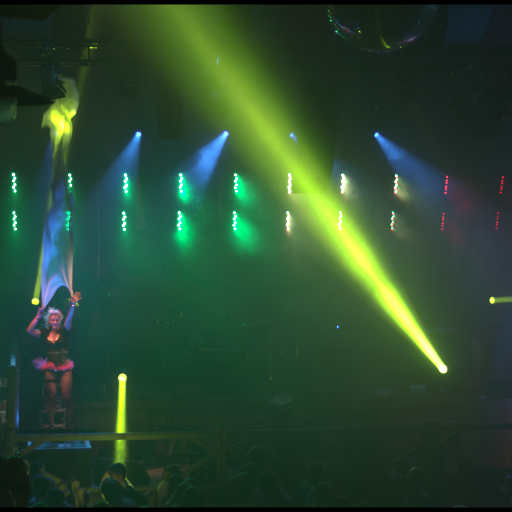}&
    \includegraphics[width=0.19\linewidth]{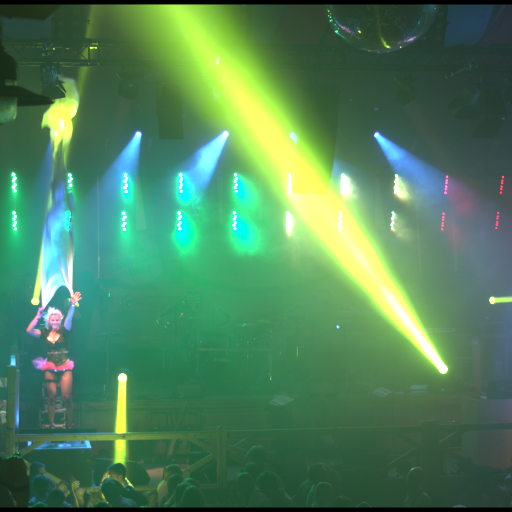}&
    \includegraphics[width=0.19\linewidth]{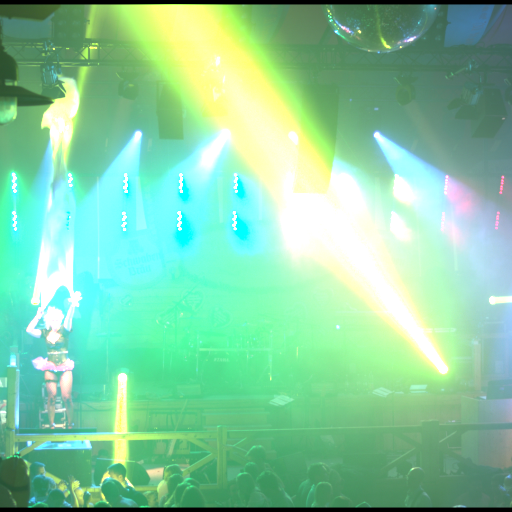}&
    \includegraphics[width=0.19\linewidth]{img/visual_beer/beerfest_lightshow_04_004202_+2.png}
    \end{tabular}    
    \caption{\small{A visual comparison of all tested methods. The frame is part of the sequence Beerfest\_04 \cite{Froehlich+2014} and shows our method has reconstruction performance similar to SAN and EIL.}}
    \label{fig:visual:beerfest}
\end{figure}
\endgroup

\subsection{Limitations}
Our network, to learn texture and dynamic range details from a single SDR video in an effective way, needs to view moving people/objects and/or view the scene from different point-of-views through camera motion. This is because over-exposed or under-exposed parts of the video may become well-exposed when these parts are not static.
When the motion in an input SDR video is limited, our network may not be able to discover how to recover texture and dynamic range in under-exposed and over-exposed parts of the video. Figure~\ref{fig:limitation} shows a frame from the sequence Bistro\_01\cite{Froehlich+2014} and the expanded frames at exposures -4~f-stops and -2~f-stops. This scene has very limited motion, only the candles in the background (green square), that limits the recovery capabilities of our method to only the flames of the candles.

\begingroup
\setlength\tabcolsep{0pt}
\renewcommand{\arraystretch}{0}
\begin{figure*}
    \centering
    \begin{tabular}{cccccc}
    &\textbf{-4 fstop} & \textbf{-2 fstop} & \textbf{0 fstop} & \textbf{+2 fstop}& \textbf{+4 fstop}\\
    \raisebox{1.5\height}{\rotatebox[origin=c]{90}{OUR}\ }&
    \includegraphics[width=0.18\linewidth]{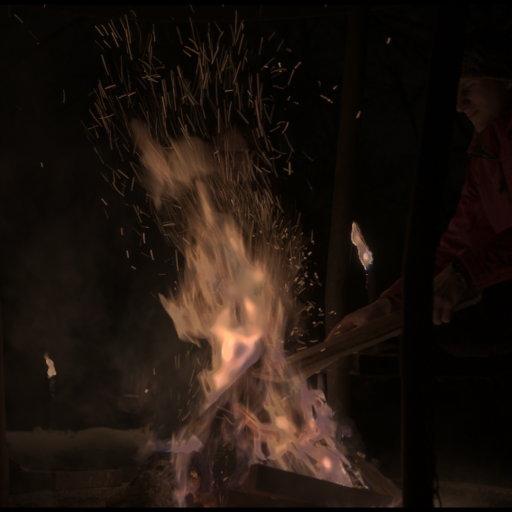}&
    \includegraphics[width=0.18\linewidth]{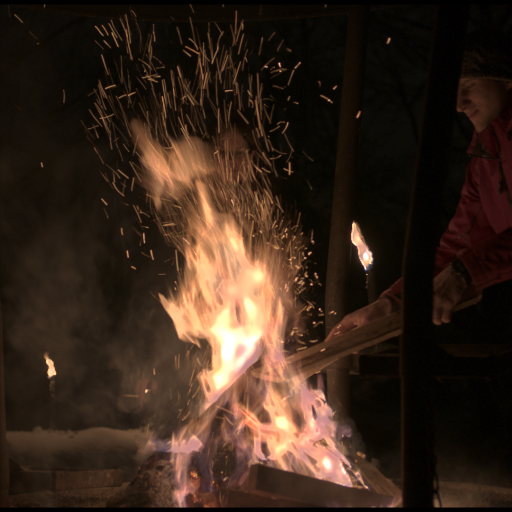}&
    \includegraphics[width=0.18\linewidth]{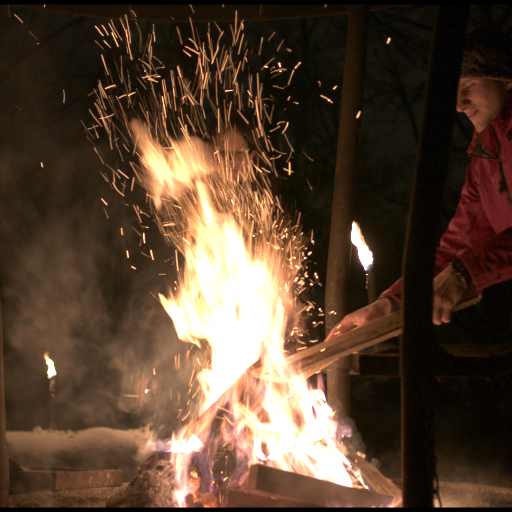}&
    \includegraphics[width=0.18\linewidth]{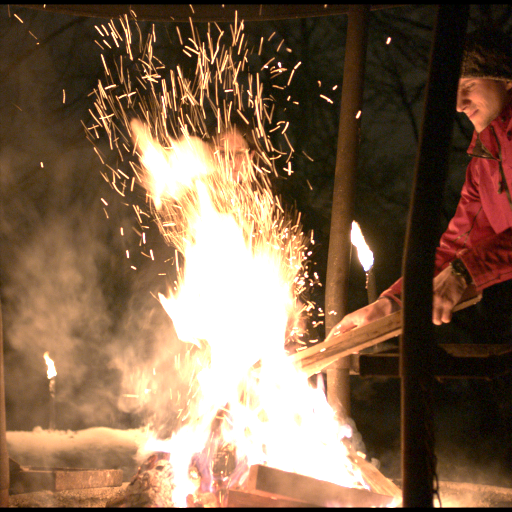}&
    \includegraphics[width=0.18\linewidth]{img/visual_fireplace/fireplace_02_000580_our_+2.png}\\
    \raisebox{1.5\height}{\rotatebox[origin=c]{90}{SAN
    }}&
    \includegraphics[width=0.18\linewidth]{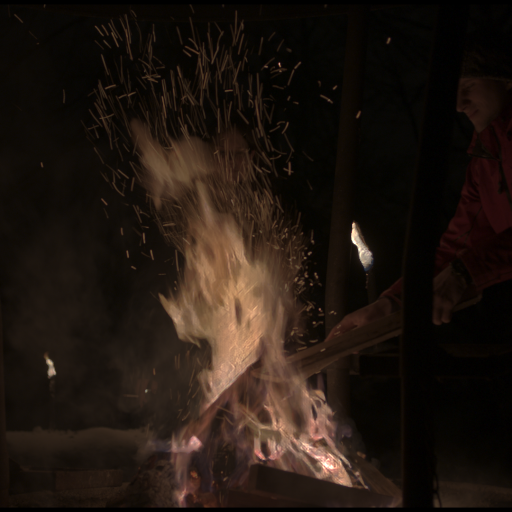}&
    \includegraphics[width=0.18\linewidth]{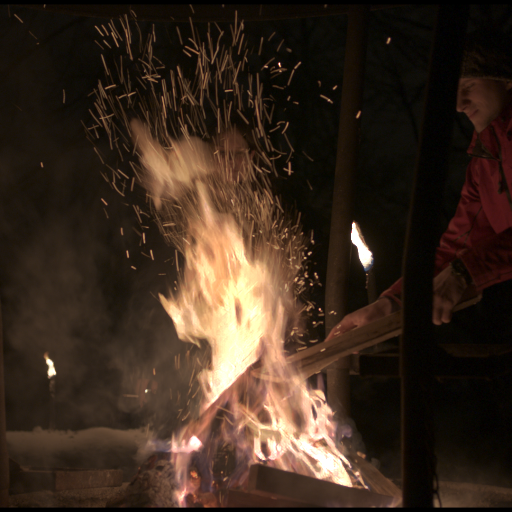}&
    \includegraphics[width=0.18\linewidth]{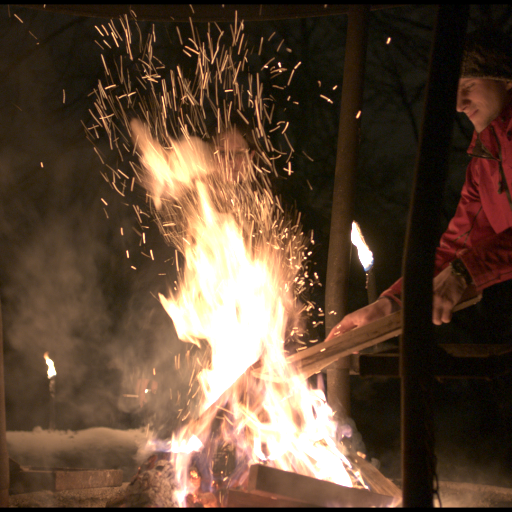}&
    \includegraphics[width=0.18\linewidth]{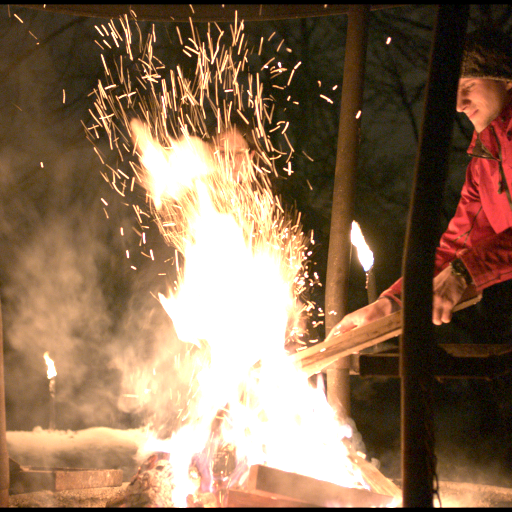}&
    \includegraphics[width=0.18\linewidth]{img/visual_fireplace/fireplace_02_000580_s_+2.png}\\
    \raisebox{1.5\height}{\rotatebox[origin=c]{90}{EIL
    }}&
    \includegraphics[width=0.18\linewidth]{img/visual_fireplace/fireplace_02_000580_s_-4.png}&
    \includegraphics[width=0.18\linewidth]{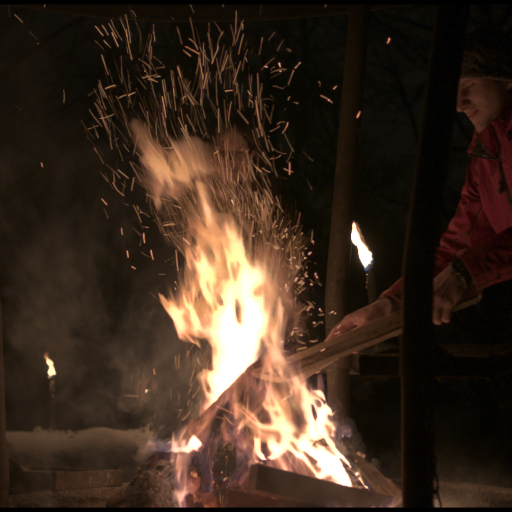}&
    \includegraphics[width=0.18\linewidth]{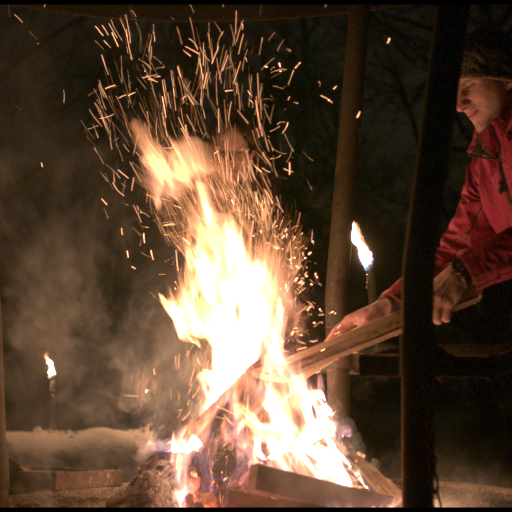}&
    \includegraphics[width=0.18\linewidth]{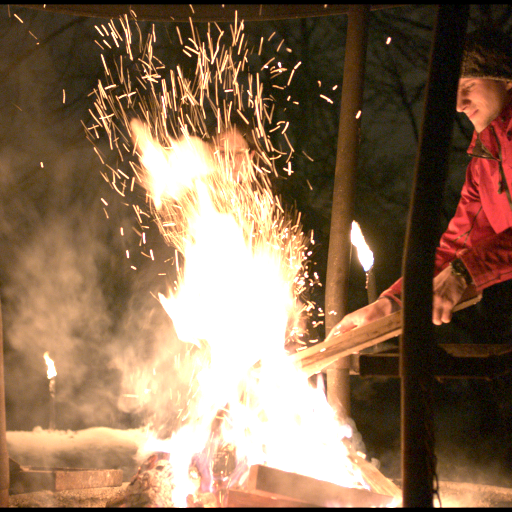}&
    \includegraphics[width=0.18\linewidth]{img/visual_fireplace/fireplace_02_000580_e_+2.png}\\
    \raisebox{1.5\height}{\rotatebox[origin=c]{90}{EXP
    }}&
    \includegraphics[width=0.18\linewidth]{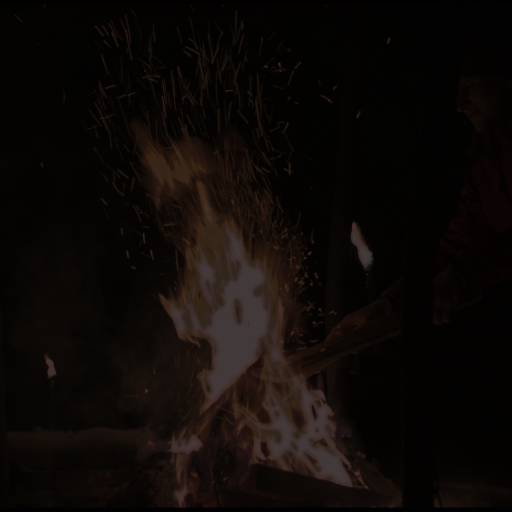}&
    \includegraphics[width=0.18\linewidth]{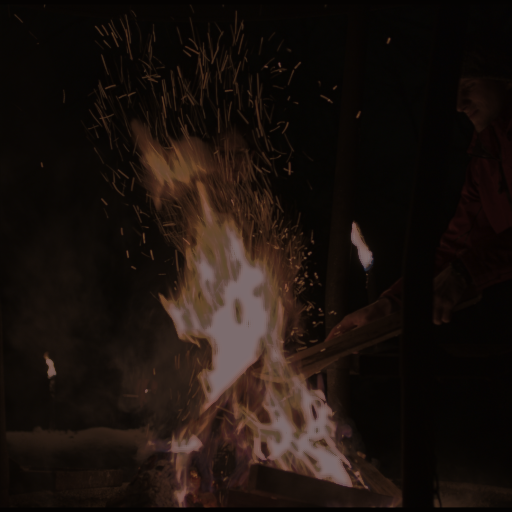}&
    \includegraphics[width=0.18\linewidth]{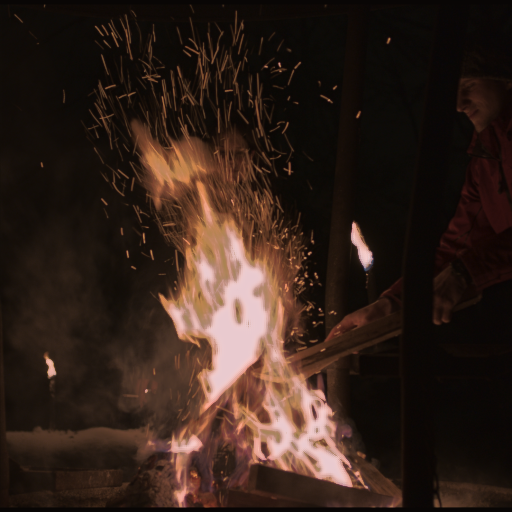}&
    \includegraphics[width=0.18\linewidth]{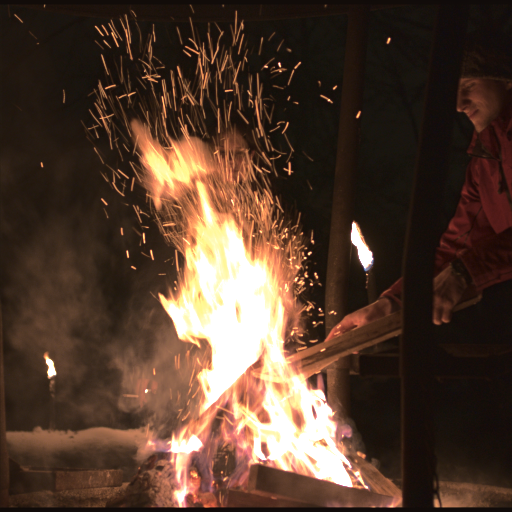}&
    \includegraphics[width=0.18\linewidth]{img/visual_fireplace/fireplace_02_000580_en_+2.png}\\
    \raisebox{1.5\height}{\rotatebox[origin=c]{90}{DRT
    }}&
    \includegraphics[width=0.18\linewidth]{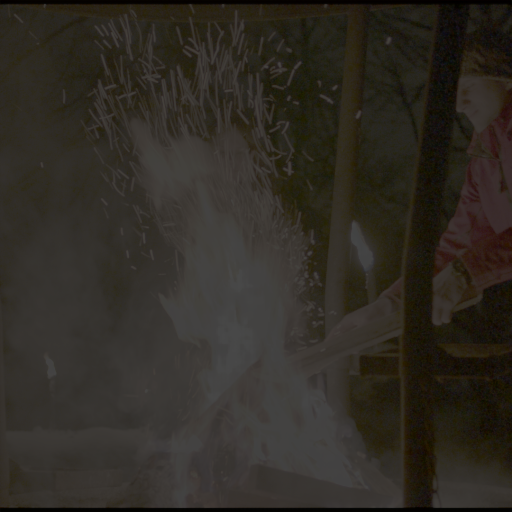}&
    \includegraphics[width=0.18\linewidth]{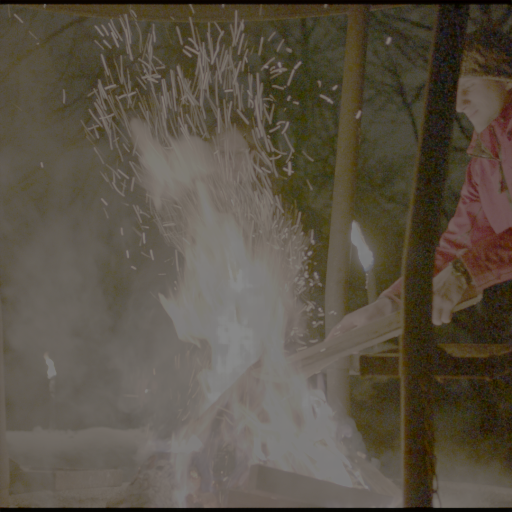}&
    \includegraphics[width=0.18\linewidth]{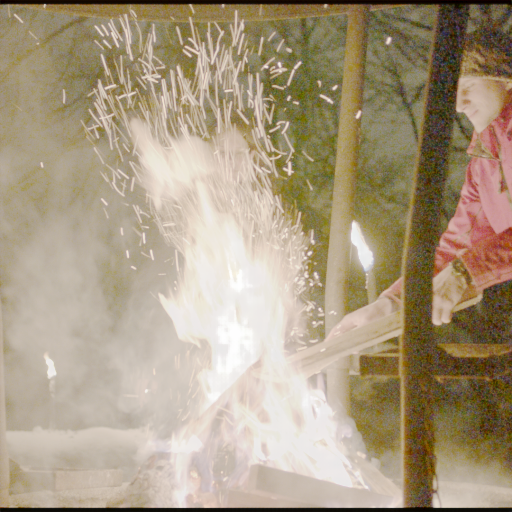}&
    \includegraphics[width=0.18\linewidth]{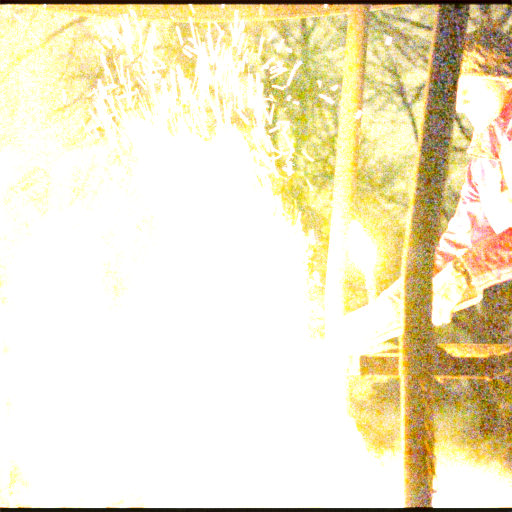}&
    \includegraphics[width=0.18\linewidth]{img/visual_fireplace/fireplace_02_000580_dr_+2.png}\\
    \raisebox{2\height}{\rotatebox[origin=c]{90}{GT
    }}&
    \includegraphics[width=0.18\linewidth]{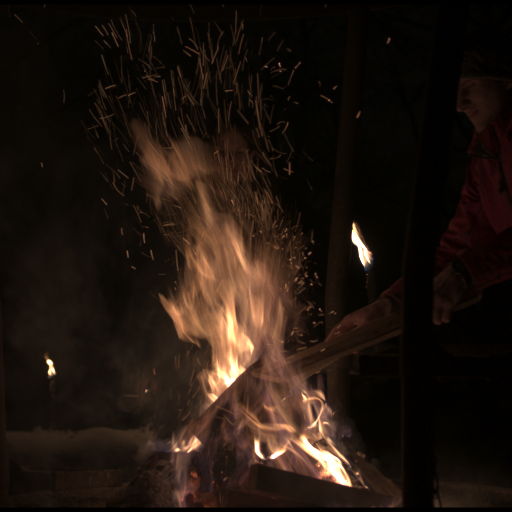}&
    \includegraphics[width=0.18\linewidth]{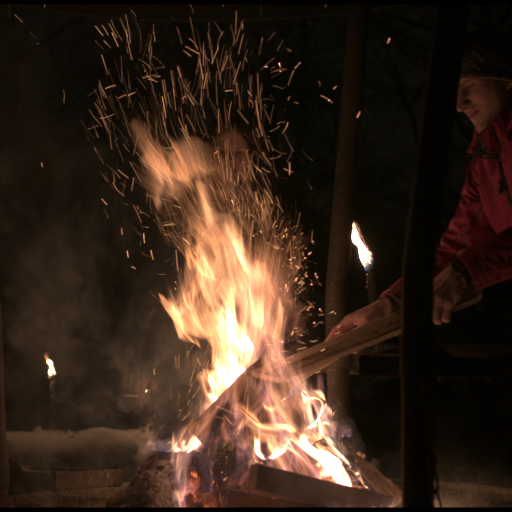}&
    \includegraphics[width=0.18\linewidth]{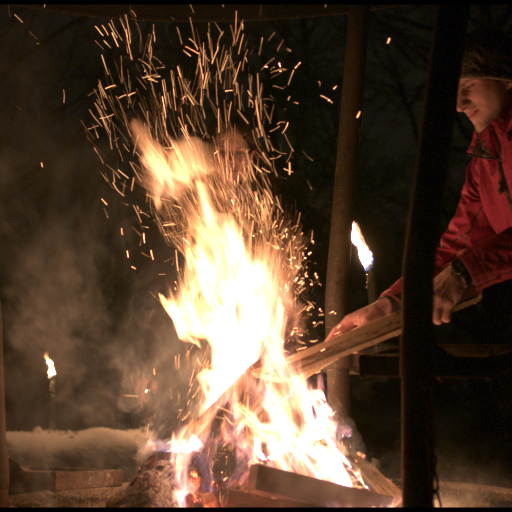}&
    \includegraphics[width=0.18\linewidth]{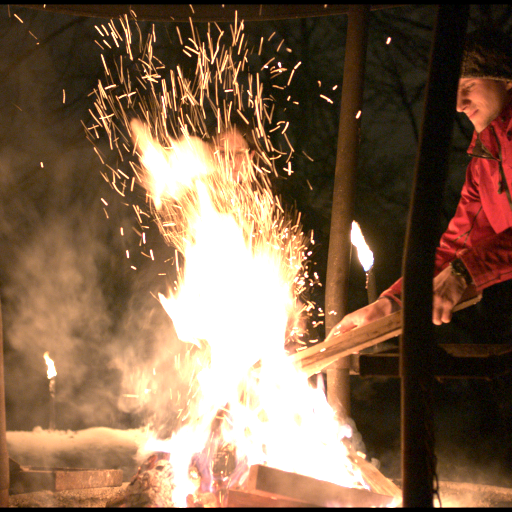}&
    \includegraphics[width=0.18\linewidth]{img/visual_fireplace/fireplace_02_000580_+2.png}
    \end{tabular}    
    \caption{\small{A visual comparison of all tested methods. The frame is part of the sequence Fireplace\_02 \cite{Froehlich+2014}. Different exposures of the resulting HDR image are shown from left to right, and different ITMOs are shown vertically.}}
    \label{fig:visual:fireplace}
\end{figure*}
\endgroup

\section{Conclusions and Future Work}
In this work, we have shown that an unsupervised approach can expand the dynamic range of SDR videos and it is possible to recover both missing details in terms of texture and dynamic range. To achieve this, we have employed zero-shot strategies. 
The proposed method can achieve high-quality results that improve on fully-supervised state-of-the-art techniques both visually and in terms of several metrics. This is particularly useful as it does not require reliance upon an external HDR dataset. 
The method performs best when there is motion in the video; ideally both in terms of people/objects and camera motion that exhibit different exposures across the frames such that the training process can form a fuller understanding of the scene's dynamic range.
This work confirms our hypothesis that SDR videos can be expanded without an external dataset and produce reasonably high-quality results that are competitive with fully-supervised methods. In future work, we would like to generalize our method and apply it to existing ITMOs for fine-tuning to provide temporal coherency and optimize training weights to the content of the input video. 



\section*{Acknowledgements}
We thank Jan Fr\"{o}hlich and his team for the Stuttgart HDR Video dataset, and Panos Nasiopoulos and his team for the UBC HDR Video dataset.

{\small
\bibliographystyle{ieee_fullname}
\bibliography{main.bib}
}

\end{document}